\documentclass[aps,prl,twocolumn]{revtex4-2}
\usepackage{amsmath,bm}
\usepackage{graphicx,epsfig,braket,amssymb,amsfonts,eufrak,tabularx,multirow,color}
\usepackage{newtxtext}
\usepackage[colorlinks=true]{hyperref}
\usepackage[T1]{fontenc}
\usepackage{textcomp}

\newcommand{\bd}{\begin{displaymath}}
\newcommand{\ed}{\end{displaymath}}
\renewcommand{\v}[1]{{\bm #1}}
\newcommand{\bpm}{\begin{pmatrix}}
\newcommand{\epm}{\end{pmatrix}}
\newcommand{\nn}{\nonumber \\}

\begin{document}

\title{Helicity-Selective Phonon Absorption and Phonon-Induced Spin Torque\\from Interfacial Spin-Lattice Coupling}

\author{Gyungchoon Go}
\email{gyungchoon@gmail.com}
\affiliation{Department of Physics, Korea Advanced Institute of Science and Technology, Daejeon 34141, Korea}

\author{Se Kwon Kim}
\email{sekwonkim@kaist.ac.kr}
\affiliation{Department of Physics, Korea Advanced Institute of Science and Technology, Daejeon 34141, Korea}

\begin{abstract}
In magnetic heterostructures with broken inversion symmetry, the Rashba effect gives rise to a gradient-free interaction
between magnons and phonons, which we term interfacial spin-lattice coupling.
Here, we investigate the dynamic consequences of this interfacial coupling in ferromagnetic heterostructures.
By expressing the interaction in terms of circular variables for magnetization and lattice displacement,
we reveal a direct interface-induced helicity-helicity coupling hat does not rely on lattice deformation gradients.
Consequently, it leads to helicity-dependent phonon absorption,
enabling in-plane acoustic waves to exert a spin torque on the magnetization, which becomes dominant in thin magnetic films.
Our findings highlight the crucial, yet overlooked, role of inversion-asymmetric interfaces in angular-momentum conversion between spin and lattice, opening up possibilities for efficient phonon-driven magnetic devices that are enabled by interface engineering.
\end{abstract}

\maketitle

\emph{Introduction}\textemdash
The interaction between spin and lattice degrees of freedom constitutes a central theme in modern condensed-matter physics.
In conventional ferromagnets, this coupling is primarily mediated by magnetostriction, where gradient of lattice deformation (strain) modifies the magnetic anisotropy~\cite{Kittel1949}.
This interaction has been widely utilized to drive strain-induced magnetization switching~\cite{Zheng2004, Bihler2008, Weiler2009, Davis2010, Roy2011, Biswas2017, Begue2021} and realize the mutual interconversion between acoustic waves and magnetization dynamics~\cite{Bommel1959, Pomerantz1961, Ganguly1976, Weiler2011, Dreher2012, Weiler2012, Sasaki2017, Tateno2020, Puebla2020,Zhang2020,Yu2020,An2020,Cai2023}.
In contrast to the symmetric strain tensor governing conventional magnetoelasticity,
the antisymmetric part of the displacement gradient, known as local lattice rotation,
couples to the magnetization $\v {m}$ in the form $\v {m} \cdot (\nabla \times \dot{\v{u}})$~\cite{Matsuo2013, Matsuo2017},
where $\v u$ is lattice displacement.
This interaction represents a universal coupling mechanism between magnetization and displacement in bulk magnetic materials,
existing independently of spin-orbit coupling or specific crystal symmetries.
It has been shown to drive acoustic spin pumping~\cite{Matsuo2013, Matsuo2017, Hamada2015, Funato2022} and facilitate angular momentum conversion between magnons and phonons~\cite{Funato2024}.

Magnetic systems possess interfaces either to the vacuum or to the other systems, which are known to give rise to novel magnetic phenomena unique to interfaces that do not exist in the bulk~\cite{Hellman2017}. However, the previous efforts on the interfacial magnetism have been largely focused on magnetic and electric sectors, leaving elastic sectors behind. Filling this gap, it has recently been shown that the presence of a physical interface gives rise to a gradient-free spin-lattice coupling~\cite{Go2026} in contrast to the aforementioned bulk spin-lattice interaction that requires finite gradient of lattice displacement.
More specifically, magnetic heterostructures with broken mirror symmetry are shown to host
${\cal L}_{\text{SL}} \propto \v{m} \cdot (\hat{\v{z}} \times \dot{\v{u}})$~\cite{Go2026},
where $\hat{\v{z}}$ denotes the interfacial normal.
This interface-induced spin-lattice coupling is explicitly gradient-free and,
therefore, can dominate the bulk counterpart relying on a finite displacement gradient in thin-film magnetic structures.
While it was shown to generate nonreciprocal phonon transport~\cite{Go2026},
its implications for angular-momentum transfer between magnons and phonons, as well as the resulting magnetization dynamics remain unexplored.

\begin{figure}[t]
\includegraphics[width=1.0\columnwidth]{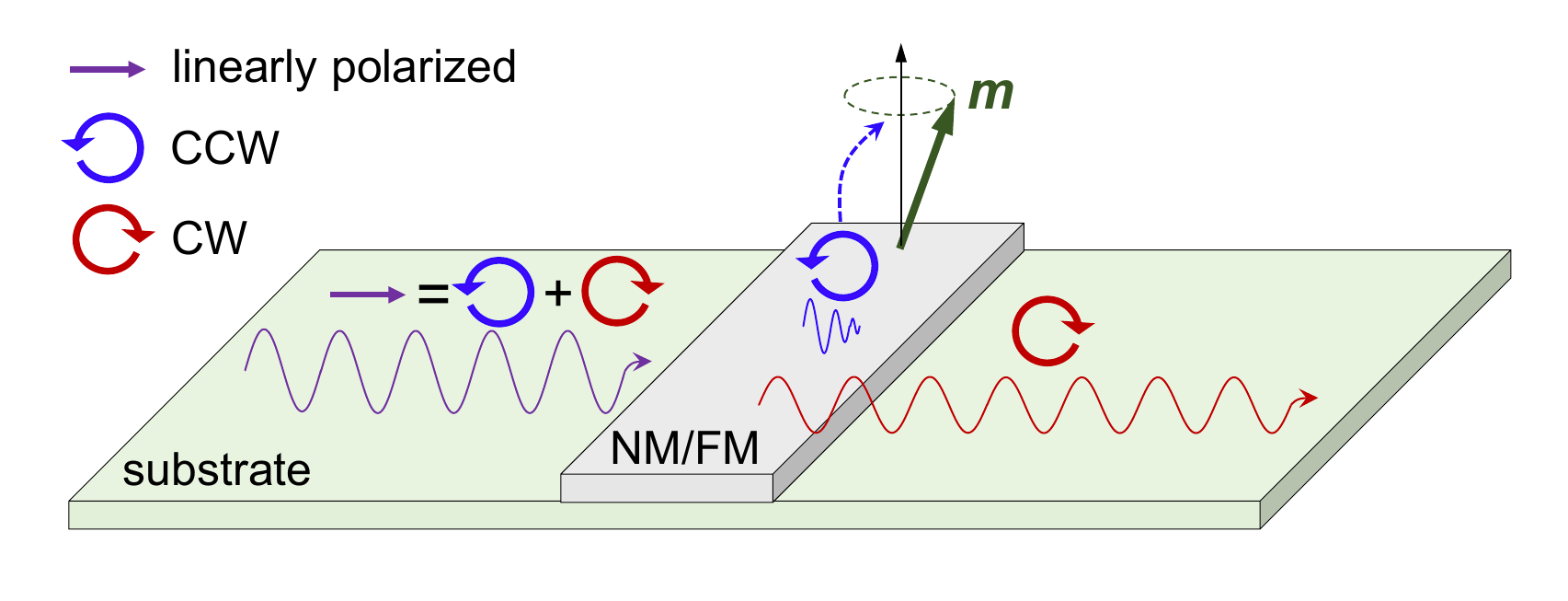}
\caption{\textbf{Schematic illustration of helicity-dependent phonon absorption and phonon-induced spin torque.}
A linearly polarized acoustic wave propagating on a substrate can be decomposed into CCW (blue) and CW (red) circular components.
Upon interacting with the normal metal/ferromagnet (NM/FM) bilayer, the CCW phonon resonantly couples to the magnetization precession and is absorbed, thereby exerting a spin torque.
In stark contrast, the off-resonant CW phonon propagates freely through the heterostructure without absorption, realizing a magnetization-controlled helicity filter for phonons.
}\label{fig:1}
\end{figure}

In this Letter, we investigate how the interfacial spin-lattice coupling enables angular-momentum transfer
and modifies magnetization dynamics in inversion-asymmetric magnetic heterostructures.
To make the underlying structure transparent, we introduce circular variables representing the magnon and phonon modes of definite helicities:
$m_\pm = (m_x \pm i m_y)/\sqrt{2}$ and $u_\pm = (u_x \pm i u_y)/\sqrt{2}$.
For a ferromagnet magnetized along the $z$ direction, the interfacial coupling becomes
\begin{equation}\label{SLcirc}
\mathcal{L}_{\text{SL}}
=
-i\lambda_{\text{SL}}
\left(
m_- \dot u_+ - m_+ \dot u_-
\right),
\end{equation}
where $\lambda_\text{SL}=n{\cal P}\alpha_{R}(m_\text{eff}-m_\text{e})/\hbar$ is the interfacial coupling constant determined by the Rashba spin-orbit interaction, with $n$ the number density of magnetic atoms, $\mathcal{P}$ the spin polarization, $\alpha_{R}$ the Rashba parameter, and $m_\text{eff}-m_\text{e}$ the difference between the effective and bare electron masses. This structure yields two helicity-resolved magnon-phonon coupling channels that are explicitly gradient-free.
Consequently, it enables an efficient angular-momentum exchange whose strength is proportional to the Rashba spin-orbit interaction
and thus can be significant in magnetic thin films.

In a ferromagnet, magnetization dynamics is intrinsically chiral and favors a single helicity determined by the equilibrium magnetization direction.
The specific helicity of the magnetization dynamics combined with the helicity-resolved coupling between magnons and phonons makes the magnets act as a helicity filter for phonons
so that, among the two helicity-resolved coupling channels, only one is efficiently activated,
leading to helicity-dependent phonon absorption.
As schematically illustrated in Fig.~\ref{fig:1}, this helicity-selective coupling results in an acoustic propagation length that depends strongly on the phonon helicity.
Specifically, when a linearly polarized acoustic wave—a superposition of clockwise (CW) and counter-clockwise (CCW) circular components—is injected, the resonant component is selectively absorbed to drive magnetization precession, while the off-resonant component propagates freely.
The resulting magnetization dynamics injects a pure spin current into the adjacent normal metal~\cite{Tserkovnyak2002prl,Tserkovnyak2002prb,Tserkovnyak2005, Saitoh2006, Costache2006, Ando2008, Czeschka2011, Morota2011}.
We find that the angular dependence of this spin current is distinct from that of previously reported acoustic spin-pumping signals driven by the bulk magnetostrictive mechanism~\cite{Dreher2012, Weiler2012}, serving as a unique fingerprint of the interfacial coupling, while its magnitude remains comparable to these conventional signals.
Our findings showcase that the interface-induced gradient-free coupling enables efficient transfer of angular momentum between lattice motion and magnetization,
revealing the overlooked effects of interfaces in the magnetization and the lattice dynamics.

\emph{Model}\textemdash
We consider a ferromagnetic thin film in the $xy$-plane on a substrate, which breaks inversion symmetry.
The equilibrium magnetization is taken along the $z$-axis $({\v m}_0 = \hat {\v z})$ in the main text for the concrete discussions.
The general cases with arbitrary equilibrium magnetization direction are discussed in the Supplemental Material~\cite{SM}.
Small transverse fluctuations are parameterized as
$\delta \v m \approx (m_x,m_y,0)$ with $|m_{x,y}|\ll1$ and the in-plane lattice displacement is denoted by $\v u=(u_x,u_y)$.

The quadratic Lagrangian density of the system is given by $\mathcal{L} = \mathcal{L}_{\text{m}} + \mathcal{L}_{\text{ph}} + \mathcal{L}_{\text{SL}}$:
\begin{equation}\label{L0}
\begin{aligned}
\mathcal{L} =& \underbrace{\frac{\rho_s}{2} \epsilon_{ij} m_i \dot{m}_j - \frac{1}{2}\left[\Delta (m_i)^2 + A_{\text{ex}} (\nabla m_i)^2\right]}_{\mathcal{L}_{\text{m}}} \\
&+\underbrace{\frac{\rho}{2} \dot{\v{u}}^2 - \frac{1}{2} \v{u} \cdot \v D \cdot \v{u} }_{\mathcal{L}_{\text{ph}}}
-\underbrace{ i\lambda_{\text{SL}}
\left(m_- \dot u_+ - m_+ \dot u_- \right)}_{\mathcal{L}_{\text{SL}}}
\end{aligned}
\end{equation}
where repeated indices $i,j\in\{x,y\}$ are summed.
Here $\rho$ is the mass density and
${\v D}_{ij}=(\lambda_L+\mu)\partial_i\partial_j+\mu\nabla^2\delta_{ij}$
is the isotropic elastic operator with Lamé coefficients $(\lambda_L,\mu)$.
In $\mathcal{L}_{\text{m}}$, $\rho_s$ denotes the spin density,
$\epsilon_{ij}$ the two-dimensional Levi–Civita tensor,
$\Delta=\gamma\rho_s B_{\rm eff}$ the magnon gap,
and $A_{\text{ex}}$ the exchange stiffness.
The term $\mathcal{L}_{\text{SL}}$ captures the interfacial spin-lattice interaction introduced in Eq.~\eqref{SLcirc}.
For $\v{m}_0=\hat{\v{z}}$ and in-plane phonons, the conventional single-ion magnetoelastic interaction~\cite{Kittel1949} yields no linear coupling,
as it vanishes at quadratic order when neglecting the out-of-plane displacement gradient--a valid assumption in the thin-film structures.

\begin{figure}[t]
\includegraphics[width=1.0\columnwidth]{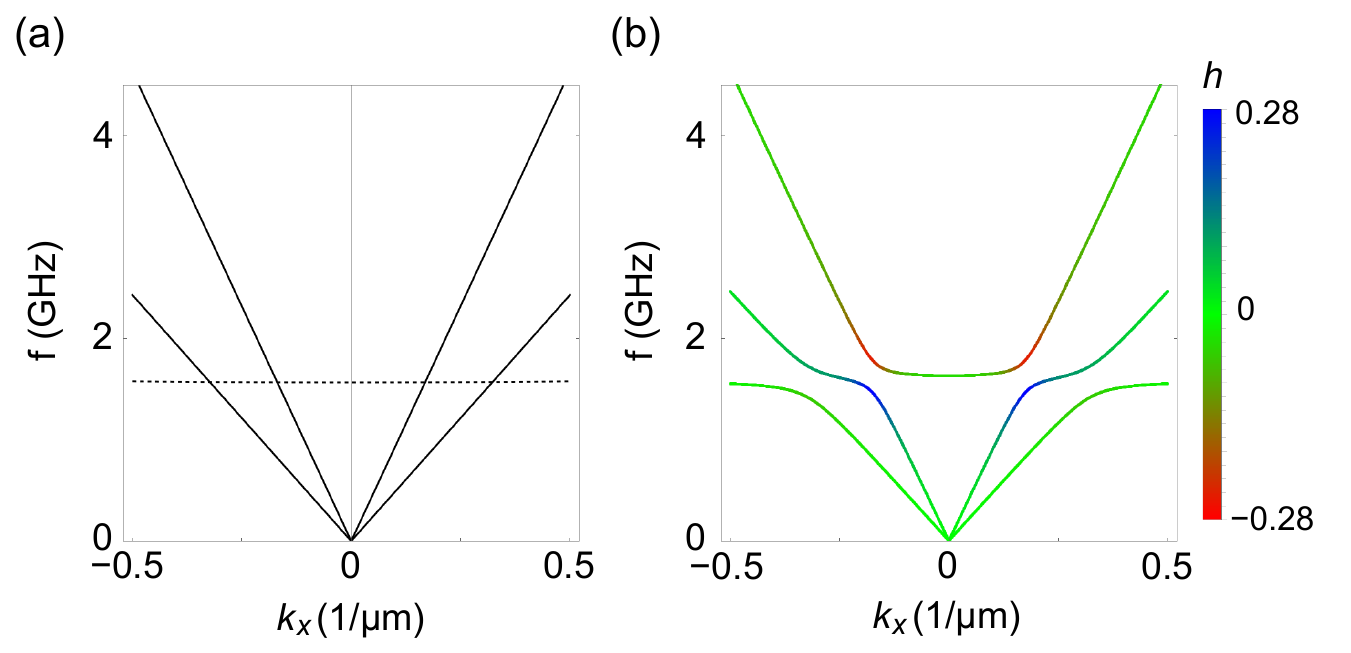}
\caption{\textbf{Magnon–phonon hybrid dispersion.}
(a) Dispersion relations of magnons and phonons along $k_x$ ($k_y = 0$) in the absence of interfacial spin-lattice coupling ($\lambda_{SL}=0$). The dashed line represents the magnon mode, while the solid lines denote the linear dispersions of the longitudinal and transverse phonon modes.
(b) Hybridized dispersions in the presence of interfacial spin-lattice coupling.
The color scale illustrates the phonon helicity weighted by the phonon energy ratio: $h = \frac{|u_+|^2 - |u_-|^2}{|u_+|^2 + |u_-|^2} \frac{E_\text{ph}}{E_\text{ph} + E_\text{mag}}$
, where $E_\text{ph}$ ($E_\text{mag}$) denotes the bare phonon (magnon) energy. The results are obtained for an effective magnetic field of $B_{\text{eff}}=0.056$~T.
}\label{fig:2}
\end{figure}

As shown in Eq.~\eqref{SLcirc}, the interfacial coupling $\mathcal{L}_{\text{SL}}$ directly pairs the magnon and phonon helicities through two distinct conjugate branches.
Remarkably, this helicity-helicity coupling is strictly gradient-free, distinguishing it from previously reported mechanisms based on magnetoelastic~\cite{An2020} or spin-vorticity couplings~\cite{Funato2024} that inherently require spatial gradients of lattice deformations.
This helicity-resolved structure governs the hybrid magnon-phonon dispersion along $k_x$ ($k_y = 0$) shown in Fig.~\ref{fig:2}(b)
and ensures angular momentum exchange between the two subsystems.
In particular, hybridization transfers helicity from magnons to phonons, thereby generating a finite phonon helicity.

Throughout this study, we adopt representative parameters appropriate for interfacial ferromagnets such as Pt/Co.
For the magnetic sector, we take a spin density $\rho_s = 7.2 \times 10^{28}\,\hbar/\mathrm{m}^3$ and an exchange stiffness $A_{\text{ex}} = 2.1 \times 10^{-11}\,\mathrm{J/m}$~\cite{Eyrich2012}.
The elastic parameters are $\rho = 8.8\,\mathrm{g/cm}^3$, $\mu = 82\,\mathrm{GPa}$ and $\lambda_L = 135\,\mathrm{GPa}$~\cite{Mcskimin1955}.
For the interfacial electronic contribution, we use $m_{\text{eff}} = 1.5 m_e$, $\alpha_R = 2\,\mathrm{eV\AA}$, and ${\cal P} = 0.5$~\cite{Miron2010}.

\emph{Helicity-dependent phonon propagation}\textemdash
The interfacial spin-lattice coupling in Eq.~\eqref{SLcirc} consists of two helicity-resolved coupling channels.
Because ferromagnetic precession selects a single helicity,
only one channel is efficiently activated, leading to helicity-dependent phonon absorption.
Including the Gilbert damping $\alpha$, the magnetization dynamics obeys the Landau-Lifshitz-Gilbert equation.
In momentum space, the equation of motion reads
\begin{equation}\label{eqLLG}
\begin{aligned}
\rho_s \dot{\v m}
=-\v m \times {\boldsymbol\Omega_{\v k}} + \alpha \rho_s \v m \times \dot{\v m} +
\boldsymbol\tau_\text{SL},
\end{aligned}
\end{equation}
where $\boldsymbol\Omega_{\v k} = -A k^2 {\v m}_\perp + \Delta \hat {\v z}$
collects the exchange and uniform precession terms.
The last term,
\begin{equation}\label{eqtSL}
\begin{aligned}
\boldsymbol\tau_{\text{SL}} = -\gamma\rho_s \v m\times {\v B}_\text{SL}
=\lambda_{\text{SL}}\,\v m\times\!\bigl(\hat {\v z} \times \dot{\v u}\bigr).
\end{aligned}
\end{equation}
represents the phonon-induced spin torque generated by the interfacial spin-lattice coupling.

\begin{figure}[t]
\includegraphics[width=1.0\columnwidth]{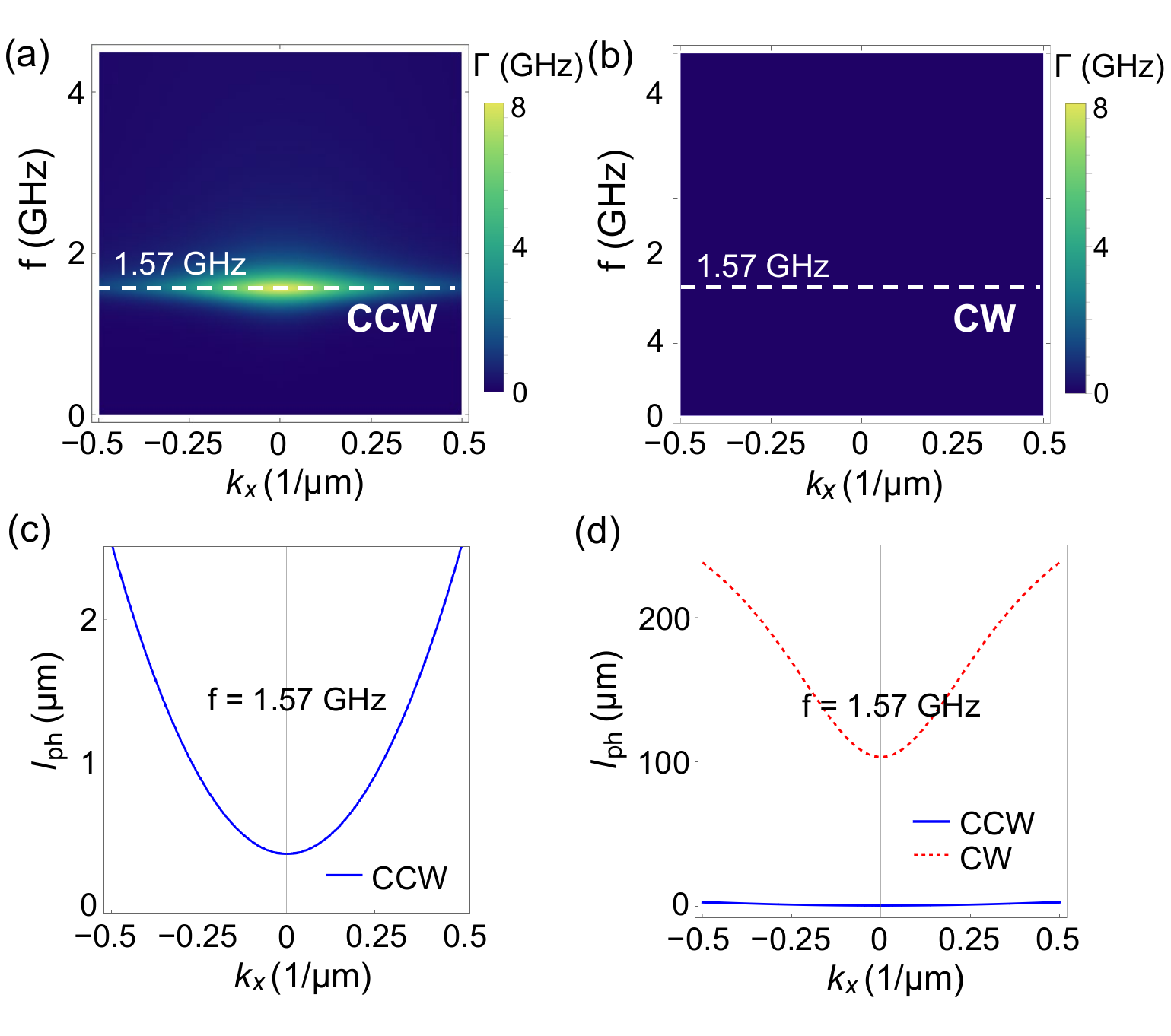}
\caption{\textbf{Helicity-dependent phonon absorption and propagation length.}
(a,b) Phonon absorption rates, $\Gamma_\text{SL}=\Delta P/\langle \mathcal{E}_\text{ph}\rangle$, for (a) CCW and (b) CW helicities, with the transferred power from phonons to magnons $\Delta P$ [Eq.~\eqref{DeltaP}].
(c,d) Corresponding phonon propagation lengths, $l_{\text{ph}}\approx v_g/(\Gamma_0+\Gamma_\text{SL})$,
with $v_g$ the group velocity, $\Gamma_0=\omega/Q$ the intrinsic phonon damping rate and $Q$ the phonon quality factor.
The parameters used are $B_{\text{eff}}=0.056$~T, $\alpha=0.1$, and $Q=1000$.
}\label{fig:3}
\end{figure}

To evaluate the phonon absorption, we express the magnetization response to the phonon-induced effective field $\v m = \boldsymbol\chi {\v B}_{\text{SL}}$.
The dynamic susceptibility $\boldsymbol\chi$ is diagonal in the helicity basis, such that $m_\pm = \chi_\pm {B}_{\text{SL}}^\pm$,
where
\begin{equation}
\chi_\pm
=
\frac{\gamma}
{\tilde{\Omega}_{\v k} \pm\omega  - i \alpha \omega},
\qquad
\tilde{\Omega}_{\v k} = \frac{A k^2 + \Delta}{\rho_s},
\end{equation}
and $B_{\text{SL}}^\pm = (B_{\text{SL}}^x \pm i B_{\text{SL}}^y)/\sqrt{2}$.
This representation directly reveals the helicity dependence of the magnetization response.
The power transferred from phonons to magnons is given by~\cite{Dreher2012,Tateno2020}
\begin{align}\label{DeltaP}
\Delta P = \frac{\gamma \omega \rho_s}{2} {\text{Im}}\left[{\v B}^\ast_{\text{SL}}(\omega, \v k)  \boldsymbol\chi (\omega, \v k){\v B}_{\text{SL}}(\omega, \v k)\right].
\end{align}
We define the absorption rate, which corresponds to the inverse of the phonon lifetime, as $\Gamma_\text{SL} = {\Delta P}/{\langle {\cal E}_{\text{ph}}\rangle}$
where $\langle {\cal E}_{\text{ph}}\rangle$ is the time-averaged phonon energy.
As shown in Fig.~\ref{fig:3}(a) and (b), the phonon absorption exhibits a strong helicity dependence.
Specifically, injecting a CCW phonon ${\v u}_\text{CCW} \propto (\cos\omega t, \sin\omega t)$
produces a pronounced absorption peak at the ferromagnetic resonance ($\omega/ 2\pi \approx 1.57$ GHz).
At this peak, the absorption rate reaches 8 GHz, corresponding to a short phonon lifetime of 125 ps.
In stark contrast, a CW phonon,
${\v u}_{\text{CW}} \propto (\cos\omega t, -\sin\omega t)$,
exhibits no resonance peak; its absorption rate remains negligibly small ($\approx$ 20 MHz), corresponding to a lifetime of approximately 50 ns.
Figure~\ref{fig:3}(c) shows the corresponding phonon propagation lengths,
$l_{\text{ph}} \approx v_g/(\Gamma_0 + \Gamma_\text{SL})$,
where $v_g = \sqrt{\mu/\rho}$ is the group velocity, $\Gamma_0 = \omega/Q$ (yielding $\Gamma_0\approx 1.57$ MHz at the ferromagnetic resonance for $Q = 1000$),
and $Q$ denotes the intrinsic phonon quality factor.
A strong contrast in propagation length emerges between the two helicities.
Near the ferromagnetic resonance frequency, the CCW phonon is strongly damped,
with its propagation length reduced to $\mu$m order, whereas the CW phonon propagates over a distance of $> 100\ \mu$m,
with their roles interchanged upon reversing the magnetization direction.
By contrast, reported experimental phonon absorption arising from conventional magnetoelastic interactions typically
amounts to only a few percent over sub-millimeter propagation distances~\cite{Sasaki2017,Tateno2020}
highlighting the remarkably stronger, helicity-selective absorption expected from our theoretical model.

\emph{Phonon-induced magnetization precession}\textemdash
The interfacial spin-lattice coupling leads to energy transfer from phonons to magnons, generating a spin torque on the magnetization.
Linearizing Eqs.~\eqref{eqLLG} and \eqref{eqtSL}, we obtain
\begin{equation}\label{Lmeq}
\begin{aligned}
m_\pm(\omega,\v k)
=
{\cal A}_\pm(\omega,\v k)
\,u_\pm(\omega,\v k),
\end{aligned}
\end{equation}
where ${\cal A}_\pm(\omega,\v k)= -
{\lambda_{\rm SL} \omega}/[\rho_s({\omega \pm \tilde\Omega_{\v k} \mp i \alpha\omega})]$.
In a ferromagnet, only one magnon helicity is resonantly enhanced for a phonon drive.
In particular, the $m_-$ mode exhibits a resonance at $\omega=\tilde\Omega_{\v k}$,
whereas the opposite-helicity mode $m_+$ remains off-resonant.
Consequently, one generally finds $|{\cal A}_-| > |{\cal A}_+|$,
so that magnetization precession is predominantly driven by the $u_-$ phonon component,
whereas the $u_+$ component remains weakly coupled.
This asymmetry reflects helicity-selective angular-momentum transfer from phonons to magnons.

\begin{figure}[t]
\includegraphics[width=1.0\columnwidth]{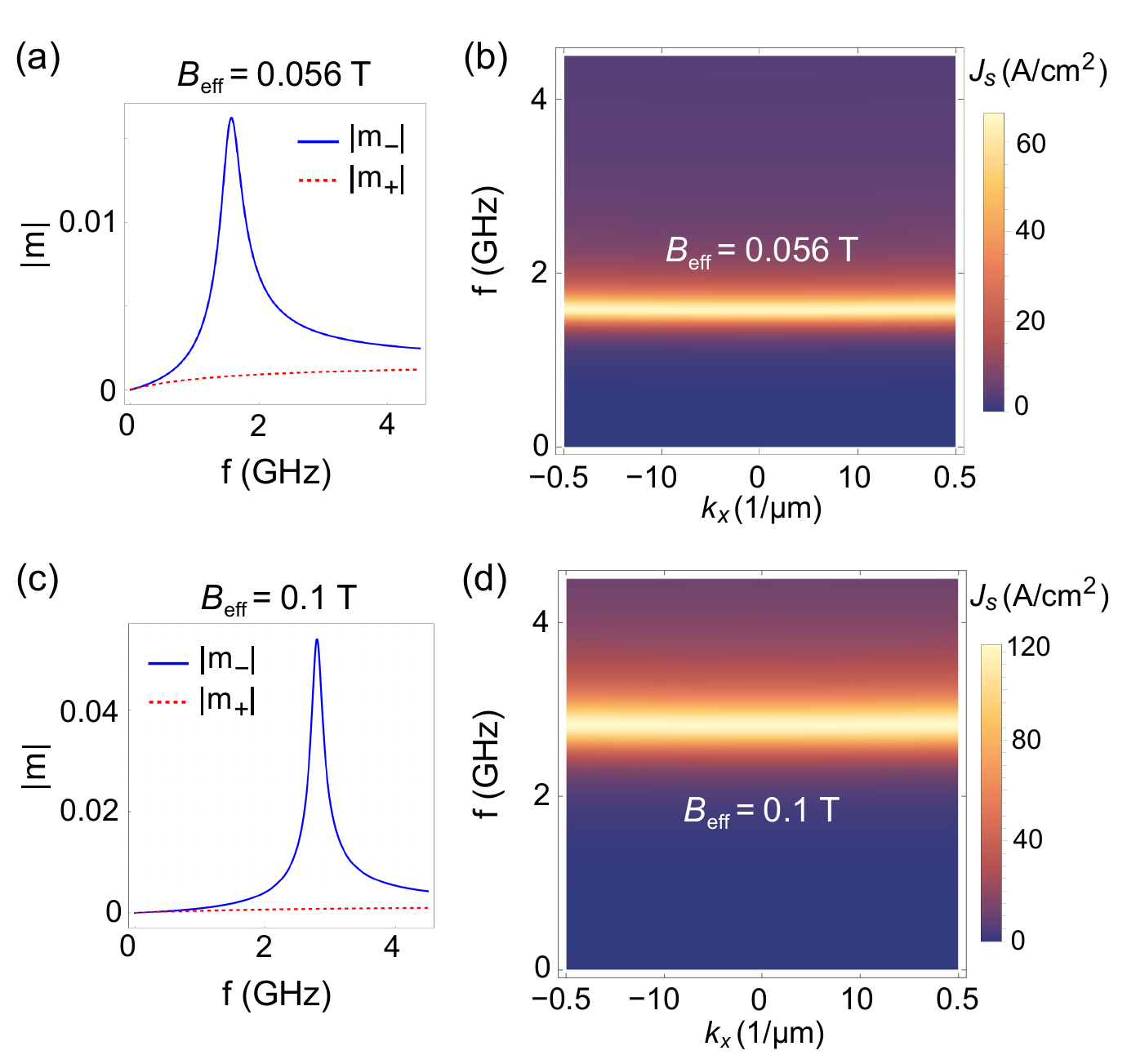}
\caption{\textbf{Phonon-driven magnetization precession and ensuing spin pumping.}
(a, c) Amplitudes of magnetization precession~[Eq.~\eqref{Lmeq}] and (b, d) the corresponding spin current profiles~[Eq.~\eqref{Jpump}].
The results are obtained for effective magnetic fields of $B_{\text{eff}} = 0.056$~T [top panels: (a, b)] and $B_{\text{eff}} = 0.1$~T [bottom panels: (c, d)].
The color map shows the pumped spin current density, expressed in electrical units as $J_s = \frac{2e}{\hbar} j_s^\text{pump}$.
The parameters used are $\alpha=0.1$ and $u_0 = 3.5$ pm.
}\label{fig:4}
\end{figure}

The intrinsic chirality of ferromagnetic precession allows for energy absorption even from a linearly polarized phonon injection.
A linearly polarized wave can be decomposed into equal-amplitude opposite circular components,
but the magnetization effectively couples to only the resonant component through the interfacial coupling.
For instance, consider a linearly polarized phonon generated by a surface acoustic wave (SAW), $u_x = u_0 \cos\omega t$ and $u_y = 0$,
corresponding to $|u_+| = |u_-| = u_0/\sqrt2$.
Although the injected phonon carries no net angular momentum,
this excitation generates a finite spin torque through the helicity-selective ferromagnetic response.
In a heterostructure with an adjacent spin sink, the resulting magnetization dynamics pumps a spin current into the neighboring normal metal  layer~\cite{Tserkovnyak2002prl,Tserkovnyak2002prb,Tserkovnyak2005, Saitoh2006, Costache2006, Ando2008, Czeschka2011,Morota2011}.
For a monochromatic steady state at frequency $\omega$, the dc $z$-component reads
\begin{equation}
\begin{aligned}
j_s^\text{pump}
&=
\frac{\hbar \omega}{8\pi}\text{Re}[g_{\uparrow\downarrow}]\left(|m_+|^2-|m_-|^2\right) \\
&= \frac{\hbar \omega}{16\pi}\text{Re}[g_{\uparrow\downarrow}]\left(|{\cal A}_+|^2-|{\cal A}_-|^2\right)u_0^2,
\end{aligned}\label{Jpump}
\end{equation}
where $g_{\uparrow\downarrow}$ is the spin-mixing conductance~\cite{Brataas2000}.

To estimate the spin current, we adopt $\text{Re}[g_{\uparrow\downarrow}] = 2 \times 10^{19}$ m$^{-2}$ and a Gilbert damping parameter of $\alpha = 0.1$
which is consistent with Ref.~\cite{Dreher2012}.
We set the lattice displacement amplitude to $u_0 = 3.5$ pm, consistent with the value used in Ref.~\cite{Funato2022}. For $\omega/2\pi = 1.57$ GHz, this corresponds to a strain $\epsilon_{xx} \approx 10^{-5}$, comparable to the strain amplitudes reported in Refs.~\cite{Weiler2011, Dreher2012, Weiler2012}.
Figure~\ref{fig:4} shows the magnetization precession amplitudes and the corresponding spin current generated by a linearly polarized phonon drive. A pronounced asymmetry between $|m_-|$ and $|m_+|$ is observed [Figs.~\ref{fig:4}(a) and (c)], resulting in a finite dc spin current [Figs.~\ref{fig:4}(b) and (d)].
Expressed in units of charge current density (by multiplying by $2e/\hbar$), the generated spin current exhibits a value comparable to previously reported acoustic spin-pumping signals under similar strain amplitudes and SAW frequencies~\cite{Dreher2012, Weiler2012}.

\emph{Discussion and outlook}\textemdash
In this work, we have shown that interfacial spin-lattice coupling gives rise to two central dynamical effects:
helicity-dependent phonon propagation and phonon-induced spin torque.
Unlike conventional magnetoelastic couplings, this interaction directly couples magnon and phonon helicities.
The intrinsic chirality of ferromagnetic precession selects a single circular component
of the phonon field, resulting in strong helicity-dependent phonon absorption and propagation.
Consequently, phonon injection can generate a finite spin torque even in the absence of net phonon angular momentum.

Observing helicity-dependent phonon propagation requires the generation of circularly polarized phonons, which can be achieved by superposing two orthogonally propagating pseudo-longitudinal SAWs~\cite{Taga2025} or by integrating materials that host chiral phonons~\cite{Zhu2018,Jeong2022, Ishito2023, Ohe2024,Kim2025}.
In contrast, the phonon-driven spin torque does not necessitate such complex setups; it can be efficiently generated even with a single linearly polarized SAW. This dynamic torque drives magnetization precession, which subsequently pumps a pure spin current into the adjacent normal metal.
Although our main text focuses on perpendicularly magnetized systems---where detecting the pumped spin current via the inverse spin Hall effect (ISHE) can be geometrically unfavorable---the phonon-driven magnetization precession induced by the interfacial coupling is not restricted to this perpendicular geometry. As detailed in the Supplemental Material~\cite{SM}, the precession persists robustly even for purely in-plane magnetizations.
Specifically, when the magnetization lies within the $xz$-plane and the phonon is polarized along the $x$-axis, the generated spin current maintains a constant maximum amplitude. This distinct angular behavior serves as a unique fingerprint of the interfacial coupling, standing in stark contrast to the strong angular variations characteristic of conventional magnetoelastic spin pumping~\cite{Dreher2012, Weiler2012}.
Consequently, spin current generated by proposed mechanism can be detected by via standard ISHE measurements~\cite{Saitoh2006, Costache2006, Ando2008, Czeschka2011, Morota2011}. Quantitative estimates for a typical Pt/FM bilayer predict an ISHE voltage signal on the order of microvolts (see Supplemental Material~\cite{SM}), which is comparable to the values reported for conventional elastic-wave-driven spin currents~\cite{Dreher2012, Weiler2012}.

We emphasize that the interfacial coupling discussed here is fundamentally distinct from bulk magnetoelastic~\cite{Weiler2011, Dreher2012, Weiler2012} and spin-vorticity~\cite{Matsuo2013, Matsuo2017} interactions.
While the latter two mechanisms are bulk effects driven by the displacement gradient $\propto \nabla \v u$,
our mechanism leverages interfacial inversion symmetry breaking to couple the spin directly to the local lattice velocity $\propto \dot{\v u}$.
As the magnetic layer thickness is reduced, the relative contribution of this interfacial pathway naturally becomes more prominent,
emerging as a primary channel for spin-mechanical conversion in heterostructures with strong Rashba spin-orbit coupling.
Our work highlights the crucial role of inversion-broken interfaces in angular-momentum transfer, broadening the fundamental understanding of spin-mechanics and providing a practical pathway for integrating phononic and spintronic technologies in thin magnetic films.

\vspace{3mm}

\emph{Acknowledgments}\textemdash
This was supported by the Brain Pool Plus Program through the National Research Foundation of Korea funded by the Ministry of Science and ICT (2020H1D3A2A03099291) and the National Research Foundation of Korea (NRF) grant funded by the Korea government (MSIT) (RS-2026-25470048).
G.G. acknowledges support by the National Research Foundation of Korea (NRF-2022R1C1C2006578).

\bibliography{reference_hel}

\begin{thebibliography}{47}%
\makeatletter
\providecommand \@ifxundefined [1]{%
 \@ifx{#1\undefined}
}%
\providecommand \@ifnum [1]{%
 \ifnum #1\expandafter \@firstoftwo
 \else \expandafter \@secondoftwo
 \fi
}%
\providecommand \@ifx [1]{%
 \ifx #1\expandafter \@firstoftwo
 \else \expandafter \@secondoftwo
 \fi
}%
\providecommand \natexlab [1]{#1}%
\providecommand \enquote  [1]{``#1''}%
\providecommand \bibnamefont  [1]{#1}%
\providecommand \bibfnamefont [1]{#1}%
\providecommand \citenamefont [1]{#1}%
\providecommand \href@noop [0]{\@secondoftwo}%
\providecommand \href [0]{\begingroup \@sanitize@url \@href}%
\providecommand \@href[1]{\@@startlink{#1}\@@href}%
\providecommand \@@href[1]{\endgroup#1\@@endlink}%
\providecommand \@sanitize@url [0]{\catcode `\\12\catcode `\$12\catcode
  `\&12\catcode `\#12\catcode `\^12\catcode `\_12\catcode `\%12\relax}%
\providecommand \@@startlink[1]{}%
\providecommand \@@endlink[0]{}%
\providecommand \url  [0]{\begingroup\@sanitize@url \@url }%
\providecommand \@url [1]{\endgroup\@href {#1}{\urlprefix }}%
\providecommand \urlprefix  [0]{URL }%
\providecommand \Eprint [0]{\href }%
\providecommand \doibase [0]{https://doi.org/}%
\providecommand \selectlanguage [0]{\@gobble}%
\providecommand \bibinfo  [0]{\@secondoftwo}%
\providecommand \bibfield  [0]{\@secondoftwo}%
\providecommand \translation [1]{[#1]}%
\providecommand \BibitemOpen [0]{}%
\providecommand \bibitemStop [0]{}%
\providecommand \bibitemNoStop [0]{.\EOS\space}%
\providecommand \EOS [0]{\spacefactor3000\relax}%
\providecommand \BibitemShut  [1]{\csname bibitem#1\endcsname}%
\let\auto@bib@innerbib\@empty
\bibitem [{\citenamefont {Kittel}(1949)}]{Kittel1949}%
  \BibitemOpen
  \bibfield  {author} {\bibinfo {author} {\bibfnamefont {C.}~\bibnamefont
  {Kittel}},\ }\bibfield  {title} {\bibinfo {title} {Physical theory of
  ferromagnetic domains},\ }\href {https://doi.org/10.1103/RevModPhys.21.541}
  {\bibfield  {journal} {\bibinfo  {journal} {Rev. Mod. Phys.}\ }\textbf
  {\bibinfo {volume} {21}},\ \bibinfo {pages} {541} (\bibinfo {year}
  {1949})}\BibitemShut {NoStop}%
\bibitem [{\citenamefont {Zheng}\ \emph {et~al.}(2004)\citenamefont {Zheng},
  \citenamefont {Wang}, \citenamefont {Lofland}, \citenamefont {Ma},
  \citenamefont {Mohaddes-Ardabili}, \citenamefont {Zhao}, \citenamefont
  {Salamanca-Riba}, \citenamefont {Shinde}, \citenamefont {Ogale},
  \citenamefont {Bai}, \citenamefont {Viehland}, \citenamefont {Jia},
  \citenamefont {Schlom}, \citenamefont {Wuttig}, \citenamefont {Roytburd},\
  and\ \citenamefont {Ramesh}}]{Zheng2004}%
  \BibitemOpen
  \bibfield  {author} {\bibinfo {author} {\bibfnamefont {H.}~\bibnamefont
  {Zheng}}, \bibinfo {author} {\bibfnamefont {J.}~\bibnamefont {Wang}},
  \bibinfo {author} {\bibfnamefont {S.~E.}\ \bibnamefont {Lofland}}, \bibinfo
  {author} {\bibfnamefont {Z.}~\bibnamefont {Ma}}, \bibinfo {author}
  {\bibfnamefont {L.}~\bibnamefont {Mohaddes-Ardabili}}, \bibinfo {author}
  {\bibfnamefont {T.}~\bibnamefont {Zhao}}, \bibinfo {author} {\bibfnamefont
  {L.}~\bibnamefont {Salamanca-Riba}}, \bibinfo {author} {\bibfnamefont
  {S.~R.}\ \bibnamefont {Shinde}}, \bibinfo {author} {\bibfnamefont {S.~B.}\
  \bibnamefont {Ogale}}, \bibinfo {author} {\bibfnamefont {F.}~\bibnamefont
  {Bai}}, \bibinfo {author} {\bibfnamefont {D.}~\bibnamefont {Viehland}},
  \bibinfo {author} {\bibfnamefont {Y.}~\bibnamefont {Jia}}, \bibinfo {author}
  {\bibfnamefont {D.~G.}\ \bibnamefont {Schlom}}, \bibinfo {author}
  {\bibfnamefont {M.}~\bibnamefont {Wuttig}}, \bibinfo {author} {\bibfnamefont
  {A.}~\bibnamefont {Roytburd}},\ and\ \bibinfo {author} {\bibfnamefont
  {R.}~\bibnamefont {Ramesh}},\ }\bibfield  {title} {\bibinfo {title}
  {{Multiferroic BaTiO$_3$-CoFe$_2$O$_4$ Nanostructures}},\ }\href
  {https://doi.org/10.1126/science.1094207} {\bibfield  {journal} {\bibinfo
  {journal} {Science}\ }\textbf {\bibinfo {volume} {303}},\ \bibinfo {pages}
  {661} (\bibinfo {year} {2004})}\BibitemShut {NoStop}%
\bibitem [{\citenamefont {Bihler}\ \emph {et~al.}(2008)\citenamefont {Bihler},
  \citenamefont {Althammer}, \citenamefont {Brandlmaier}, \citenamefont
  {Gepr\"ags}, \citenamefont {Weiler}, \citenamefont {Opel}, \citenamefont
  {Schoch}, \citenamefont {Limmer}, \citenamefont {Gross}, \citenamefont
  {Brandt},\ and\ \citenamefont {Goennenwein}}]{Bihler2008}%
  \BibitemOpen
  \bibfield  {author} {\bibinfo {author} {\bibfnamefont {C.}~\bibnamefont
  {Bihler}}, \bibinfo {author} {\bibfnamefont {M.}~\bibnamefont {Althammer}},
  \bibinfo {author} {\bibfnamefont {A.}~\bibnamefont {Brandlmaier}}, \bibinfo
  {author} {\bibfnamefont {S.}~\bibnamefont {Gepr\"ags}}, \bibinfo {author}
  {\bibfnamefont {M.}~\bibnamefont {Weiler}}, \bibinfo {author} {\bibfnamefont
  {M.}~\bibnamefont {Opel}}, \bibinfo {author} {\bibfnamefont {W.}~\bibnamefont
  {Schoch}}, \bibinfo {author} {\bibfnamefont {W.}~\bibnamefont {Limmer}},
  \bibinfo {author} {\bibfnamefont {R.}~\bibnamefont {Gross}}, \bibinfo
  {author} {\bibfnamefont {M.~S.}\ \bibnamefont {Brandt}},\ and\ \bibinfo
  {author} {\bibfnamefont {S.~T.~B.}\ \bibnamefont {Goennenwein}},\ }\bibfield
  {title} {\bibinfo {title} {{Ga$_{1-x}$Mn$_{x}$As$/$piezoelectric actuator
  hybrids: A model system for magnetoelastic magnetization manipulation}},\
  }\href {https://doi.org/10.1103/PhysRevB.78.045203} {\bibfield  {journal}
  {\bibinfo  {journal} {Phys. Rev. B}\ }\textbf {\bibinfo {volume} {78}},\
  \bibinfo {pages} {045203} (\bibinfo {year} {2008})}\BibitemShut {NoStop}%
\bibitem [{\citenamefont {Weiler}\ \emph {et~al.}(2009)\citenamefont {Weiler},
  \citenamefont {Brandlmaier}, \citenamefont {Geprägs}, \citenamefont
  {Althammer}, \citenamefont {Opel}, \citenamefont {Bihler}, \citenamefont
  {Huebl}, \citenamefont {Brandt}, \citenamefont {Gross},\ and\ \citenamefont
  {Goennenwein}}]{Weiler2009}%
  \BibitemOpen
  \bibfield  {author} {\bibinfo {author} {\bibfnamefont {M.}~\bibnamefont
  {Weiler}}, \bibinfo {author} {\bibfnamefont {A.}~\bibnamefont {Brandlmaier}},
  \bibinfo {author} {\bibfnamefont {S.}~\bibnamefont {Geprägs}}, \bibinfo
  {author} {\bibfnamefont {M.}~\bibnamefont {Althammer}}, \bibinfo {author}
  {\bibfnamefont {M.}~\bibnamefont {Opel}}, \bibinfo {author} {\bibfnamefont
  {C.}~\bibnamefont {Bihler}}, \bibinfo {author} {\bibfnamefont
  {H.}~\bibnamefont {Huebl}}, \bibinfo {author} {\bibfnamefont {M.~S.}\
  \bibnamefont {Brandt}}, \bibinfo {author} {\bibfnamefont {R.}~\bibnamefont
  {Gross}},\ and\ \bibinfo {author} {\bibfnamefont {S.~T.~B.}\ \bibnamefont
  {Goennenwein}},\ }\bibfield  {title} {\bibinfo {title} {Voltage controlled
  inversion of magnetic anisotropy in a ferromagnetic thin film at room
  temperature},\ }\href {https://doi.org/10.1088/1367-2630/11/1/013021}
  {\bibfield  {journal} {\bibinfo  {journal} {New J. Phys.}\ }\textbf {\bibinfo
  {volume} {11}},\ \bibinfo {pages} {013021} (\bibinfo {year}
  {2009})}\BibitemShut {NoStop}%
\bibitem [{\citenamefont {Davis}\ \emph {et~al.}(2010)\citenamefont {Davis},
  \citenamefont {Baruth},\ and\ \citenamefont {Adenwalla}}]{Davis2010}%
  \BibitemOpen
  \bibfield  {author} {\bibinfo {author} {\bibfnamefont {S.}~\bibnamefont
  {Davis}}, \bibinfo {author} {\bibfnamefont {A.}~\bibnamefont {Baruth}},\ and\
  \bibinfo {author} {\bibfnamefont {S.}~\bibnamefont {Adenwalla}},\ }\bibfield
  {title} {\bibinfo {title} {Magnetization dynamics triggered by surface
  acoustic waves},\ }\href {https://doi.org/10.1063/1.3521289} {\bibfield
  {journal} {\bibinfo  {journal} {Appl. Phys. Lett.}\ }\textbf {\bibinfo
  {volume} {97}},\ \bibinfo {pages} {232507} (\bibinfo {year}
  {2010})}\BibitemShut {NoStop}%
\bibitem [{\citenamefont {Roy}\ \emph {et~al.}(2011)\citenamefont {Roy},
  \citenamefont {Bandyopadhyay},\ and\ \citenamefont {Atulasimha}}]{Roy2011}%
  \BibitemOpen
  \bibfield  {author} {\bibinfo {author} {\bibfnamefont {K.}~\bibnamefont
  {Roy}}, \bibinfo {author} {\bibfnamefont {S.}~\bibnamefont {Bandyopadhyay}},\
  and\ \bibinfo {author} {\bibfnamefont {J.}~\bibnamefont {Atulasimha}},\
  }\bibfield  {title} {\bibinfo {title} {Switching dynamics of a
  magnetostrictive single-domain nanomagnet subjected to stress},\ }\href
  {https://doi.org/10.1103/PhysRevB.83.224412} {\bibfield  {journal} {\bibinfo
  {journal} {Phys. Rev. B}\ }\textbf {\bibinfo {volume} {83}},\ \bibinfo
  {pages} {224412} (\bibinfo {year} {2011})}\BibitemShut {NoStop}%
\bibitem [{\citenamefont {Biswas}\ \emph {et~al.}(2017)\citenamefont {Biswas},
  \citenamefont {Ahmad}, \citenamefont {Atulasimha},\ and\ \citenamefont
  {Bandyopadhyay}}]{Biswas2017}%
  \BibitemOpen
  \bibfield  {author} {\bibinfo {author} {\bibfnamefont {A.~K.}\ \bibnamefont
  {Biswas}}, \bibinfo {author} {\bibfnamefont {H.}~\bibnamefont {Ahmad}},
  \bibinfo {author} {\bibfnamefont {J.}~\bibnamefont {Atulasimha}},\ and\
  \bibinfo {author} {\bibfnamefont {S.}~\bibnamefont {Bandyopadhyay}},\
  }\bibfield  {title} {\bibinfo {title} {Experimental {Demonstration} of
  {Complete} 180$^\circ$ {Reversal} of {Magnetization} in {Isolated} {Co}
  {Nanomagnets} on a {PMN}–{PT} {Substrate} with {Voltage} {Generated}
  {Strain}},\ }\href {https://doi.org/10.1021/acs.nanolett.7b00439} {\bibfield
  {journal} {\bibinfo  {journal} {Nano Lett.}\ }\textbf {\bibinfo {volume}
  {17}},\ \bibinfo {pages} {3478} (\bibinfo {year} {2017})}\BibitemShut
  {NoStop}%
\bibitem [{\citenamefont {Begué}\ and\ \citenamefont
  {Ciria}(2021)}]{Begue2021}%
  \BibitemOpen
  \bibfield  {author} {\bibinfo {author} {\bibfnamefont {A.}~\bibnamefont
  {Begué}}\ and\ \bibinfo {author} {\bibfnamefont {M.}~\bibnamefont {Ciria}},\
  }\bibfield  {title} {\bibinfo {title} {Strain-{Mediated} {Giant}
  {Magnetoelectric} {Coupling} in a {Crystalline} {Multiferroic}
  {Heterostructure}},\ }\href {https://doi.org/10.1021/acsami.0c18777}
  {\bibfield  {journal} {\bibinfo  {journal} {ACS Appl. Mater. Interfaces}\
  }\textbf {\bibinfo {volume} {13}},\ \bibinfo {pages} {6778} (\bibinfo {year}
  {2021})}\BibitemShut {NoStop}%
\bibitem [{\citenamefont {B\"ommel}\ and\ \citenamefont
  {Dransfeld}(1959)}]{Bommel1959}%
  \BibitemOpen
  \bibfield  {author} {\bibinfo {author} {\bibfnamefont {H.}~\bibnamefont
  {B\"ommel}}\ and\ \bibinfo {author} {\bibfnamefont {K.}~\bibnamefont
  {Dransfeld}},\ }\bibfield  {title} {\bibinfo {title} {Excitation of
  hypersonic waves by ferromagnetic resonance},\ }\href
  {https://doi.org/10.1103/PhysRevLett.3.83} {\bibfield  {journal} {\bibinfo
  {journal} {Phys. Rev. Lett.}\ }\textbf {\bibinfo {volume} {3}},\ \bibinfo
  {pages} {83} (\bibinfo {year} {1959})}\BibitemShut {NoStop}%
\bibitem [{\citenamefont {Pomerantz}(1961)}]{Pomerantz1961}%
  \BibitemOpen
  \bibfield  {author} {\bibinfo {author} {\bibfnamefont {M.}~\bibnamefont
  {Pomerantz}},\ }\bibfield  {title} {\bibinfo {title} {Excitation of spin-wave
  resonance by microwave phonons},\ }\href
  {https://doi.org/10.1103/PhysRevLett.7.312} {\bibfield  {journal} {\bibinfo
  {journal} {Phys. Rev. Lett.}\ }\textbf {\bibinfo {volume} {7}},\ \bibinfo
  {pages} {312} (\bibinfo {year} {1961})}\BibitemShut {NoStop}%
\bibitem [{\citenamefont {Ganguly}\ \emph {et~al.}(1976)\citenamefont
  {Ganguly}, \citenamefont {Davis}, \citenamefont {Webb},\ and\ \citenamefont
  {Vittoria}}]{Ganguly1976}%
  \BibitemOpen
  \bibfield  {author} {\bibinfo {author} {\bibfnamefont {A.~K.}\ \bibnamefont
  {Ganguly}}, \bibinfo {author} {\bibfnamefont {K.~L.}\ \bibnamefont {Davis}},
  \bibinfo {author} {\bibfnamefont {D.~C.}\ \bibnamefont {Webb}},\ and\
  \bibinfo {author} {\bibfnamefont {C.}~\bibnamefont {Vittoria}},\ }\bibfield
  {title} {\bibinfo {title} {Magnetoelastic surface waves in a magnetic
  film–piezoelectric substrate configuration},\ }\href
  {https://doi.org/10.1063/1.322991} {\bibfield  {journal} {\bibinfo  {journal}
  {J. Appl. Phys.}\ }\textbf {\bibinfo {volume} {47}},\ \bibinfo {pages} {2696}
  (\bibinfo {year} {1976})}\BibitemShut {NoStop}%
\bibitem [{\citenamefont {Weiler}\ \emph {et~al.}(2011)\citenamefont {Weiler},
  \citenamefont {Dreher}, \citenamefont {Heeg}, \citenamefont {Huebl},
  \citenamefont {Gross}, \citenamefont {Brandt},\ and\ \citenamefont
  {Goennenwein}}]{Weiler2011}%
  \BibitemOpen
  \bibfield  {author} {\bibinfo {author} {\bibfnamefont {M.}~\bibnamefont
  {Weiler}}, \bibinfo {author} {\bibfnamefont {L.}~\bibnamefont {Dreher}},
  \bibinfo {author} {\bibfnamefont {C.}~\bibnamefont {Heeg}}, \bibinfo {author}
  {\bibfnamefont {H.}~\bibnamefont {Huebl}}, \bibinfo {author} {\bibfnamefont
  {R.}~\bibnamefont {Gross}}, \bibinfo {author} {\bibfnamefont {M.~S.}\
  \bibnamefont {Brandt}},\ and\ \bibinfo {author} {\bibfnamefont {S.~T.~B.}\
  \bibnamefont {Goennenwein}},\ }\bibfield  {title} {\bibinfo {title}
  {Elastically driven ferromagnetic resonance in nickel thin films},\ }\href
  {https://doi.org/10.1103/PhysRevLett.106.117601} {\bibfield  {journal}
  {\bibinfo  {journal} {Phys. Rev. Lett.}\ }\textbf {\bibinfo {volume} {106}},\
  \bibinfo {pages} {117601} (\bibinfo {year} {2011})}\BibitemShut {NoStop}%
\bibitem [{\citenamefont {Dreher}\ \emph {et~al.}(2012)\citenamefont {Dreher},
  \citenamefont {Weiler}, \citenamefont {Pernpeintner}, \citenamefont {Huebl},
  \citenamefont {Gross}, \citenamefont {Brandt},\ and\ \citenamefont
  {Goennenwein}}]{Dreher2012}%
  \BibitemOpen
  \bibfield  {author} {\bibinfo {author} {\bibfnamefont {L.}~\bibnamefont
  {Dreher}}, \bibinfo {author} {\bibfnamefont {M.}~\bibnamefont {Weiler}},
  \bibinfo {author} {\bibfnamefont {M.}~\bibnamefont {Pernpeintner}}, \bibinfo
  {author} {\bibfnamefont {H.}~\bibnamefont {Huebl}}, \bibinfo {author}
  {\bibfnamefont {R.}~\bibnamefont {Gross}}, \bibinfo {author} {\bibfnamefont
  {M.~S.}\ \bibnamefont {Brandt}},\ and\ \bibinfo {author} {\bibfnamefont
  {S.~T.~B.}\ \bibnamefont {Goennenwein}},\ }\bibfield  {title} {\bibinfo
  {title} {Surface acoustic wave driven ferromagnetic resonance in nickel thin
  films: Theory and experiment},\ }\href
  {https://doi.org/10.1103/PhysRevB.86.134415} {\bibfield  {journal} {\bibinfo
  {journal} {Phys. Rev. B}\ }\textbf {\bibinfo {volume} {86}},\ \bibinfo
  {pages} {134415} (\bibinfo {year} {2012})}\BibitemShut {NoStop}%
\bibitem [{\citenamefont {Weiler}\ \emph {et~al.}(2012)\citenamefont {Weiler},
  \citenamefont {Huebl}, \citenamefont {Goerg}, \citenamefont {Czeschka},
  \citenamefont {Gross},\ and\ \citenamefont {Goennenwein}}]{Weiler2012}%
  \BibitemOpen
  \bibfield  {author} {\bibinfo {author} {\bibfnamefont {M.}~\bibnamefont
  {Weiler}}, \bibinfo {author} {\bibfnamefont {H.}~\bibnamefont {Huebl}},
  \bibinfo {author} {\bibfnamefont {F.~S.}\ \bibnamefont {Goerg}}, \bibinfo
  {author} {\bibfnamefont {F.~D.}\ \bibnamefont {Czeschka}}, \bibinfo {author}
  {\bibfnamefont {R.}~\bibnamefont {Gross}},\ and\ \bibinfo {author}
  {\bibfnamefont {S.~T.~B.}\ \bibnamefont {Goennenwein}},\ }\bibfield  {title}
  {\bibinfo {title} {Spin pumping with coherent elastic waves},\ }\href
  {https://doi.org/10.1103/PhysRevLett.108.176601} {\bibfield  {journal}
  {\bibinfo  {journal} {Phys. Rev. Lett.}\ }\textbf {\bibinfo {volume} {108}},\
  \bibinfo {pages} {176601} (\bibinfo {year} {2012})}\BibitemShut {NoStop}%
\bibitem [{\citenamefont {Sasaki}\ \emph {et~al.}(2017)\citenamefont {Sasaki},
  \citenamefont {Nii}, \citenamefont {Iguchi},\ and\ \citenamefont
  {Onose}}]{Sasaki2017}%
  \BibitemOpen
  \bibfield  {author} {\bibinfo {author} {\bibfnamefont {R.}~\bibnamefont
  {Sasaki}}, \bibinfo {author} {\bibfnamefont {Y.}~\bibnamefont {Nii}},
  \bibinfo {author} {\bibfnamefont {Y.}~\bibnamefont {Iguchi}},\ and\ \bibinfo
  {author} {\bibfnamefont {Y.}~\bibnamefont {Onose}},\ }\bibfield  {title}
  {\bibinfo {title} {{Nonreciprocal propagation of surface acoustic wave in
  Ni/LiNbO$_{3}$}},\ }\href {https://doi.org/10.1103/PhysRevB.95.020407}
  {\bibfield  {journal} {\bibinfo  {journal} {Phys. Rev. B}\ }\textbf {\bibinfo
  {volume} {95}},\ \bibinfo {pages} {020407} (\bibinfo {year}
  {2017})}\BibitemShut {NoStop}%
\bibitem [{\citenamefont {Tateno}\ and\ \citenamefont
  {Nozaki}(2020)}]{Tateno2020}%
  \BibitemOpen
  \bibfield  {author} {\bibinfo {author} {\bibfnamefont {S.}~\bibnamefont
  {Tateno}}\ and\ \bibinfo {author} {\bibfnamefont {Y.}~\bibnamefont
  {Nozaki}},\ }\bibfield  {title} {\bibinfo {title} {{Highly Nonreciprocal Spin
  Waves Excited by Magnetoelastic Coupling in a Ni/Si Bilayer}},\ }\href
  {https://doi.org/10.1103/PhysRevApplied.13.034074} {\bibfield  {journal}
  {\bibinfo  {journal} {Phys. Rev. Appl.}\ }\textbf {\bibinfo {volume} {13}},\
  \bibinfo {pages} {034074} (\bibinfo {year} {2020})}\BibitemShut {NoStop}%
\bibitem [{\citenamefont {Puebla}\ \emph {et~al.}(2020)\citenamefont {Puebla},
  \citenamefont {Xu}, \citenamefont {Rana}, \citenamefont {Yamamoto},
  \citenamefont {Maekawa},\ and\ \citenamefont {Otani}}]{Puebla2020}%
  \BibitemOpen
  \bibfield  {author} {\bibinfo {author} {\bibfnamefont {J.}~\bibnamefont
  {Puebla}}, \bibinfo {author} {\bibfnamefont {M.}~\bibnamefont {Xu}}, \bibinfo
  {author} {\bibfnamefont {B.}~\bibnamefont {Rana}}, \bibinfo {author}
  {\bibfnamefont {K.}~\bibnamefont {Yamamoto}}, \bibinfo {author}
  {\bibfnamefont {S.}~\bibnamefont {Maekawa}},\ and\ \bibinfo {author}
  {\bibfnamefont {Y.}~\bibnamefont {Otani}},\ }\bibfield  {title} {\bibinfo
  {title} {Acoustic ferromagnetic resonance and spin pumping induced by surface
  acoustic waves},\ }\href {https://doi.org/10.1088/1361-6463/ab7efe}
  {\bibfield  {journal} {\bibinfo  {journal} {J. Phys. D: Appl. Phys.}\
  }\textbf {\bibinfo {volume} {53}},\ \bibinfo {pages} {264002} (\bibinfo
  {year} {2020})}\BibitemShut {NoStop}%
\bibitem [{\citenamefont {Zhang}\ \emph {et~al.}(2020)\citenamefont {Zhang},
  \citenamefont {Bauer},\ and\ \citenamefont {Yu}}]{Zhang2020}%
  \BibitemOpen
  \bibfield  {author} {\bibinfo {author} {\bibfnamefont {X.}~\bibnamefont
  {Zhang}}, \bibinfo {author} {\bibfnamefont {G.~E.~W.}\ \bibnamefont
  {Bauer}},\ and\ \bibinfo {author} {\bibfnamefont {T.}~\bibnamefont {Yu}},\
  }\bibfield  {title} {\bibinfo {title} {Unidirectional pumping of phonons by
  magnetization dynamics},\ }\href
  {https://doi.org/10.1103/PhysRevLett.125.077203} {\bibfield  {journal}
  {\bibinfo  {journal} {Phys. Rev. Lett.}\ }\textbf {\bibinfo {volume} {125}},\
  \bibinfo {pages} {077203} (\bibinfo {year} {2020})}\BibitemShut {NoStop}%
\bibitem [{\citenamefont {Yu}(2020)}]{Yu2020}%
  \BibitemOpen
  \bibfield  {author} {\bibinfo {author} {\bibfnamefont {T.}~\bibnamefont
  {Yu}},\ }\bibfield  {title} {\bibinfo {title} {Nonreciprocal surface
  magnetoelastic dynamics},\ }\href
  {https://doi.org/10.1103/PhysRevB.102.134417} {\bibfield  {journal} {\bibinfo
   {journal} {Phys. Rev. B}\ }\textbf {\bibinfo {volume} {102}},\ \bibinfo
  {pages} {134417} (\bibinfo {year} {2020})}\BibitemShut {NoStop}%
\bibitem [{\citenamefont {An}\ \emph {et~al.}(2020)\citenamefont {An},
  \citenamefont {Litvinenko}, \citenamefont {Kohno}, \citenamefont {Fuad},
  \citenamefont {Naletov}, \citenamefont {Vila}, \citenamefont {Ebels},
  \citenamefont {de~Loubens}, \citenamefont {Hurdequint}, \citenamefont
  {Beaulieu}, \citenamefont {Ben~Youssef}, \citenamefont {Vukadinovic},
  \citenamefont {Bauer}, \citenamefont {Slavin}, \citenamefont {Tiberkevich},\
  and\ \citenamefont {Klein}}]{An2020}%
  \BibitemOpen
  \bibfield  {author} {\bibinfo {author} {\bibfnamefont {K.}~\bibnamefont
  {An}}, \bibinfo {author} {\bibfnamefont {A.~N.}\ \bibnamefont {Litvinenko}},
  \bibinfo {author} {\bibfnamefont {R.}~\bibnamefont {Kohno}}, \bibinfo
  {author} {\bibfnamefont {A.~A.}\ \bibnamefont {Fuad}}, \bibinfo {author}
  {\bibfnamefont {V.~V.}\ \bibnamefont {Naletov}}, \bibinfo {author}
  {\bibfnamefont {L.}~\bibnamefont {Vila}}, \bibinfo {author} {\bibfnamefont
  {U.}~\bibnamefont {Ebels}}, \bibinfo {author} {\bibfnamefont
  {G.}~\bibnamefont {de~Loubens}}, \bibinfo {author} {\bibfnamefont
  {H.}~\bibnamefont {Hurdequint}}, \bibinfo {author} {\bibfnamefont
  {N.}~\bibnamefont {Beaulieu}}, \bibinfo {author} {\bibfnamefont
  {J.}~\bibnamefont {Ben~Youssef}}, \bibinfo {author} {\bibfnamefont
  {N.}~\bibnamefont {Vukadinovic}}, \bibinfo {author} {\bibfnamefont
  {G.~E.~W.}\ \bibnamefont {Bauer}}, \bibinfo {author} {\bibfnamefont {A.~N.}\
  \bibnamefont {Slavin}}, \bibinfo {author} {\bibfnamefont {V.~S.}\
  \bibnamefont {Tiberkevich}},\ and\ \bibinfo {author} {\bibfnamefont
  {O.}~\bibnamefont {Klein}},\ }\bibfield  {title} {\bibinfo {title} {Coherent
  long-range transfer of angular momentum between magnon kittel modes by
  phonons},\ }\href {https://doi.org/10.1103/PhysRevB.101.060407} {\bibfield
  {journal} {\bibinfo  {journal} {Phys. Rev. B}\ }\textbf {\bibinfo {volume}
  {101}},\ \bibinfo {pages} {060407} (\bibinfo {year} {2020})}\BibitemShut
  {NoStop}%
\bibitem [{\citenamefont {Cai}\ \emph {et~al.}(2023)\citenamefont {Cai},
  \citenamefont {Zhou}, \citenamefont {Yu},\ and\ \citenamefont
  {Yu}}]{Cai2023}%
  \BibitemOpen
  \bibfield  {author} {\bibinfo {author} {\bibfnamefont {C.}~\bibnamefont
  {Cai}}, \bibinfo {author} {\bibfnamefont {X.-H.}\ \bibnamefont {Zhou}},
  \bibinfo {author} {\bibfnamefont {W.}~\bibnamefont {Yu}},\ and\ \bibinfo
  {author} {\bibfnamefont {T.}~\bibnamefont {Yu}},\ }\bibfield  {title}
  {\bibinfo {title} {Acoustic frequency multiplication and pure second-harmonic
  generation of phonons by magnetic transducers},\ }\href
  {https://doi.org/10.1103/PhysRevB.107.L100410} {\bibfield  {journal}
  {\bibinfo  {journal} {Phys. Rev. B}\ }\textbf {\bibinfo {volume} {107}},\
  \bibinfo {pages} {L100410} (\bibinfo {year} {2023})}\BibitemShut {NoStop}%
\bibitem [{\citenamefont {Matsuo}\ \emph {et~al.}(2013)\citenamefont {Matsuo},
  \citenamefont {Ieda}, \citenamefont {Harii}, \citenamefont {Saitoh},\ and\
  \citenamefont {Maekawa}}]{Matsuo2013}%
  \BibitemOpen
  \bibfield  {author} {\bibinfo {author} {\bibfnamefont {M.}~\bibnamefont
  {Matsuo}}, \bibinfo {author} {\bibfnamefont {J.}~\bibnamefont {Ieda}},
  \bibinfo {author} {\bibfnamefont {K.}~\bibnamefont {Harii}}, \bibinfo
  {author} {\bibfnamefont {E.}~\bibnamefont {Saitoh}},\ and\ \bibinfo {author}
  {\bibfnamefont {S.}~\bibnamefont {Maekawa}},\ }\bibfield  {title} {\bibinfo
  {title} {Mechanical generation of spin current by spin-rotation coupling},\
  }\href {https://doi.org/10.1103/PhysRevB.87.180402} {\bibfield  {journal}
  {\bibinfo  {journal} {Phys. Rev. B}\ }\textbf {\bibinfo {volume} {87}},\
  \bibinfo {pages} {180402} (\bibinfo {year} {2013})}\BibitemShut {NoStop}%
\bibitem [{\citenamefont {Matsuo}\ \emph {et~al.}(2017)\citenamefont {Matsuo},
  \citenamefont {Saitoh},\ and\ \citenamefont {Maekawa}}]{Matsuo2017}%
  \BibitemOpen
  \bibfield  {author} {\bibinfo {author} {\bibfnamefont {M.}~\bibnamefont
  {Matsuo}}, \bibinfo {author} {\bibfnamefont {E.}~\bibnamefont {Saitoh}},\
  and\ \bibinfo {author} {\bibfnamefont {S.}~\bibnamefont {Maekawa}},\
  }\bibfield  {title} {\bibinfo {title} {Spin-mechatronics},\ }\href
  {https://doi.org/10.7566/JPSJ.86.011011} {\bibfield  {journal} {\bibinfo
  {journal} {J. Phys. Soc. Jpn.}\ }\textbf {\bibinfo {volume} {86}},\ \bibinfo
  {pages} {011011} (\bibinfo {year} {2017})}\BibitemShut {NoStop}%
\bibitem [{\citenamefont {Hamada}\ \emph {et~al.}(2015)\citenamefont {Hamada},
  \citenamefont {Yokoyama},\ and\ \citenamefont {Murakami}}]{Hamada2015}%
  \BibitemOpen
  \bibfield  {author} {\bibinfo {author} {\bibfnamefont {M.}~\bibnamefont
  {Hamada}}, \bibinfo {author} {\bibfnamefont {T.}~\bibnamefont {Yokoyama}},\
  and\ \bibinfo {author} {\bibfnamefont {S.}~\bibnamefont {Murakami}},\
  }\bibfield  {title} {\bibinfo {title} {Spin current generation and magnetic
  response in carbon nanotubes by the twisting phonon mode},\ }\href
  {https://doi.org/10.1103/PhysRevB.92.060409} {\bibfield  {journal} {\bibinfo
  {journal} {Phys. Rev. B}\ }\textbf {\bibinfo {volume} {92}},\ \bibinfo
  {pages} {060409} (\bibinfo {year} {2015})}\BibitemShut {NoStop}%
\bibitem [{\citenamefont {Funato}\ and\ \citenamefont
  {Matsuo}(2022)}]{Funato2022}%
  \BibitemOpen
  \bibfield  {author} {\bibinfo {author} {\bibfnamefont {T.}~\bibnamefont
  {Funato}}\ and\ \bibinfo {author} {\bibfnamefont {M.}~\bibnamefont
  {Matsuo}},\ }\bibfield  {title} {\bibinfo {title} {Spin elastodynamic motive
  force},\ }\href {https://doi.org/10.1103/PhysRevLett.128.077201} {\bibfield
  {journal} {\bibinfo  {journal} {Phys. Rev. Lett.}\ }\textbf {\bibinfo
  {volume} {128}},\ \bibinfo {pages} {077201} (\bibinfo {year}
  {2022})}\BibitemShut {NoStop}%
\bibitem [{\citenamefont {Funato}\ \emph {et~al.}(2024)\citenamefont {Funato},
  \citenamefont {Matsuo},\ and\ \citenamefont {Kato}}]{Funato2024}%
  \BibitemOpen
  \bibfield  {author} {\bibinfo {author} {\bibfnamefont {T.}~\bibnamefont
  {Funato}}, \bibinfo {author} {\bibfnamefont {M.}~\bibnamefont {Matsuo}},\
  and\ \bibinfo {author} {\bibfnamefont {T.}~\bibnamefont {Kato}},\ }\bibfield
  {title} {\bibinfo {title} {Chirality-induced phonon-spin conversion at an
  interface},\ }\href {https://doi.org/10.1103/PhysRevLett.132.236201}
  {\bibfield  {journal} {\bibinfo  {journal} {Phys. Rev. Lett.}\ }\textbf
  {\bibinfo {volume} {132}},\ \bibinfo {pages} {236201} (\bibinfo {year}
  {2024})}\BibitemShut {NoStop}%
\bibitem [{\citenamefont {Hellman}\ \emph {et~al.}(2017)\citenamefont
  {Hellman}, \citenamefont {Hoffmann}, \citenamefont {Tserkovnyak},
  \citenamefont {Beach}, \citenamefont {Fullerton}, \citenamefont {Leighton},
  \citenamefont {MacDonald}, \citenamefont {Ralph}, \citenamefont {Arena},
  \citenamefont {D\"urr}, \citenamefont {Fischer}, \citenamefont {Grollier},
  \citenamefont {Heremans}, \citenamefont {Jungwirth}, \citenamefont {Kimel},
  \citenamefont {Koopmans}, \citenamefont {Krivorotov}, \citenamefont {May},
  \citenamefont {Petford-Long}, \citenamefont {Rondinelli}, \citenamefont
  {Samarth}, \citenamefont {Schuller}, \citenamefont {Slavin}, \citenamefont
  {Stiles}, \citenamefont {Tchernyshyov}, \citenamefont {Thiaville},\ and\
  \citenamefont {Zink}}]{Hellman2017}%
  \BibitemOpen
  \bibfield  {author} {\bibinfo {author} {\bibfnamefont {F.}~\bibnamefont
  {Hellman}}, \bibinfo {author} {\bibfnamefont {A.}~\bibnamefont {Hoffmann}},
  \bibinfo {author} {\bibfnamefont {Y.}~\bibnamefont {Tserkovnyak}}, \bibinfo
  {author} {\bibfnamefont {G.~S.~D.}\ \bibnamefont {Beach}}, \bibinfo {author}
  {\bibfnamefont {E.~E.}\ \bibnamefont {Fullerton}}, \bibinfo {author}
  {\bibfnamefont {C.}~\bibnamefont {Leighton}}, \bibinfo {author}
  {\bibfnamefont {A.~H.}\ \bibnamefont {MacDonald}}, \bibinfo {author}
  {\bibfnamefont {D.~C.}\ \bibnamefont {Ralph}}, \bibinfo {author}
  {\bibfnamefont {D.~A.}\ \bibnamefont {Arena}}, \bibinfo {author}
  {\bibfnamefont {H.~A.}\ \bibnamefont {D\"urr}}, \bibinfo {author}
  {\bibfnamefont {P.}~\bibnamefont {Fischer}}, \bibinfo {author} {\bibfnamefont
  {J.}~\bibnamefont {Grollier}}, \bibinfo {author} {\bibfnamefont {J.~P.}\
  \bibnamefont {Heremans}}, \bibinfo {author} {\bibfnamefont {T.}~\bibnamefont
  {Jungwirth}}, \bibinfo {author} {\bibfnamefont {A.~V.}\ \bibnamefont
  {Kimel}}, \bibinfo {author} {\bibfnamefont {B.}~\bibnamefont {Koopmans}},
  \bibinfo {author} {\bibfnamefont {I.~N.}\ \bibnamefont {Krivorotov}},
  \bibinfo {author} {\bibfnamefont {S.~J.}\ \bibnamefont {May}}, \bibinfo
  {author} {\bibfnamefont {A.~K.}\ \bibnamefont {Petford-Long}}, \bibinfo
  {author} {\bibfnamefont {J.~M.}\ \bibnamefont {Rondinelli}}, \bibinfo
  {author} {\bibfnamefont {N.}~\bibnamefont {Samarth}}, \bibinfo {author}
  {\bibfnamefont {I.~K.}\ \bibnamefont {Schuller}}, \bibinfo {author}
  {\bibfnamefont {A.~N.}\ \bibnamefont {Slavin}}, \bibinfo {author}
  {\bibfnamefont {M.~D.}\ \bibnamefont {Stiles}}, \bibinfo {author}
  {\bibfnamefont {O.}~\bibnamefont {Tchernyshyov}}, \bibinfo {author}
  {\bibfnamefont {A.}~\bibnamefont {Thiaville}},\ and\ \bibinfo {author}
  {\bibfnamefont {B.~L.}\ \bibnamefont {Zink}},\ }\bibfield  {title} {\bibinfo
  {title} {Interface-induced phenomena in magnetism},\ }\href
  {https://doi.org/10.1103/RevModPhys.89.025006} {\bibfield  {journal}
  {\bibinfo  {journal} {Rev. Mod. Phys.}\ }\textbf {\bibinfo {volume} {89}},\
  \bibinfo {pages} {025006} (\bibinfo {year} {2017})}\BibitemShut {NoStop}%
\bibitem [{\citenamefont {Go}\ and\ \citenamefont {Kim}(2026)}]{Go2026}%
  \BibitemOpen
  \bibfield  {author} {\bibinfo {author} {\bibfnamefont {G.}~\bibnamefont
  {Go}}\ and\ \bibinfo {author} {\bibfnamefont {S.~K.}\ \bibnamefont {Kim}},\
  }\href@noop {} {\bibinfo {title} {{Kineo-Elasticity and Nonreciprocal Phonons
  by Rashba-induced Interfacial Spin-Lattice Coupling}}} (\bibinfo {year}
  {2026}),\ \Eprint {https://arxiv.org/abs/arXiv:2601.12656} {arXiv:2601.12656}
  \BibitemShut {NoStop}%
\bibitem [{\citenamefont {Tserkovnyak}\ \emph
  {et~al.}(2002{\natexlab{a}})\citenamefont {Tserkovnyak}, \citenamefont
  {Brataas},\ and\ \citenamefont {Bauer}}]{Tserkovnyak2002prl}%
  \BibitemOpen
  \bibfield  {author} {\bibinfo {author} {\bibfnamefont {Y.}~\bibnamefont
  {Tserkovnyak}}, \bibinfo {author} {\bibfnamefont {A.}~\bibnamefont
  {Brataas}},\ and\ \bibinfo {author} {\bibfnamefont {G.~E.~W.}\ \bibnamefont
  {Bauer}},\ }\bibfield  {title} {\bibinfo {title} {Enhanced gilbert damping in
  thin ferromagnetic films},\ }\href
  {https://doi.org/10.1103/PhysRevLett.88.117601} {\bibfield  {journal}
  {\bibinfo  {journal} {Phys. Rev. Lett.}\ }\textbf {\bibinfo {volume} {88}},\
  \bibinfo {pages} {117601} (\bibinfo {year} {2002}{\natexlab{a}})}\BibitemShut
  {NoStop}%
\bibitem [{\citenamefont {Tserkovnyak}\ \emph
  {et~al.}(2002{\natexlab{b}})\citenamefont {Tserkovnyak}, \citenamefont
  {Brataas},\ and\ \citenamefont {Bauer}}]{Tserkovnyak2002prb}%
  \BibitemOpen
  \bibfield  {author} {\bibinfo {author} {\bibfnamefont {Y.}~\bibnamefont
  {Tserkovnyak}}, \bibinfo {author} {\bibfnamefont {A.}~\bibnamefont
  {Brataas}},\ and\ \bibinfo {author} {\bibfnamefont {G.~E.~W.}\ \bibnamefont
  {Bauer}},\ }\bibfield  {title} {\bibinfo {title} {Spin pumping and
  magnetization dynamics in metallic multilayers},\ }\href
  {https://doi.org/10.1103/PhysRevB.66.224403} {\bibfield  {journal} {\bibinfo
  {journal} {Phys. Rev. B}\ }\textbf {\bibinfo {volume} {66}},\ \bibinfo
  {pages} {224403} (\bibinfo {year} {2002}{\natexlab{b}})}\BibitemShut
  {NoStop}%
\bibitem [{\citenamefont {Tserkovnyak}\ \emph {et~al.}(2005)\citenamefont
  {Tserkovnyak}, \citenamefont {Brataas}, \citenamefont {Bauer},\ and\
  \citenamefont {Halperin}}]{Tserkovnyak2005}%
  \BibitemOpen
  \bibfield  {author} {\bibinfo {author} {\bibfnamefont {Y.}~\bibnamefont
  {Tserkovnyak}}, \bibinfo {author} {\bibfnamefont {A.}~\bibnamefont
  {Brataas}}, \bibinfo {author} {\bibfnamefont {G.~E.~W.}\ \bibnamefont
  {Bauer}},\ and\ \bibinfo {author} {\bibfnamefont {B.~I.}\ \bibnamefont
  {Halperin}},\ }\bibfield  {title} {\bibinfo {title} {Nonlocal magnetization
  dynamics in ferromagnetic heterostructures},\ }\href
  {https://doi.org/10.1103/RevModPhys.77.1375} {\bibfield  {journal} {\bibinfo
  {journal} {Rev. Mod. Phys.}\ }\textbf {\bibinfo {volume} {77}},\ \bibinfo
  {pages} {1375} (\bibinfo {year} {2005})}\BibitemShut {NoStop}%
\bibitem [{\citenamefont {Saitoh}\ \emph {et~al.}(2006)\citenamefont {Saitoh},
  \citenamefont {Ueda}, \citenamefont {Miyajima},\ and\ \citenamefont
  {Tatara}}]{Saitoh2006}%
  \BibitemOpen
  \bibfield  {author} {\bibinfo {author} {\bibfnamefont {E.}~\bibnamefont
  {Saitoh}}, \bibinfo {author} {\bibfnamefont {M.}~\bibnamefont {Ueda}},
  \bibinfo {author} {\bibfnamefont {H.}~\bibnamefont {Miyajima}},\ and\
  \bibinfo {author} {\bibfnamefont {G.}~\bibnamefont {Tatara}},\ }\bibfield
  {title} {\bibinfo {title} {Conversion of spin current into charge current at
  room temperature: {Inverse} spin-{Hall} effect},\ }\href
  {https://doi.org/10.1063/1.2199473} {\bibfield  {journal} {\bibinfo
  {journal} {Appl. Phys. Lett.}\ }\textbf {\bibinfo {volume} {88}},\ \bibinfo
  {pages} {182509} (\bibinfo {year} {2006})}\BibitemShut {NoStop}%
\bibitem [{\citenamefont {Costache}\ \emph {et~al.}(2006)\citenamefont
  {Costache}, \citenamefont {Sladkov}, \citenamefont {Watts}, \citenamefont
  {van~der Wal},\ and\ \citenamefont {van Wees}}]{Costache2006}%
  \BibitemOpen
  \bibfield  {author} {\bibinfo {author} {\bibfnamefont {M.~V.}\ \bibnamefont
  {Costache}}, \bibinfo {author} {\bibfnamefont {M.}~\bibnamefont {Sladkov}},
  \bibinfo {author} {\bibfnamefont {S.~M.}\ \bibnamefont {Watts}}, \bibinfo
  {author} {\bibfnamefont {C.~H.}\ \bibnamefont {van~der Wal}},\ and\ \bibinfo
  {author} {\bibfnamefont {B.~J.}\ \bibnamefont {van Wees}},\ }\bibfield
  {title} {\bibinfo {title} {Electrical detection of spin pumping due to the
  precessing magnetization of a single ferromagnet},\ }\href
  {https://doi.org/10.1103/PhysRevLett.97.216603} {\bibfield  {journal}
  {\bibinfo  {journal} {Phys. Rev. Lett.}\ }\textbf {\bibinfo {volume} {97}},\
  \bibinfo {pages} {216603} (\bibinfo {year} {2006})}\BibitemShut {NoStop}%
\bibitem [{\citenamefont {Ando}\ \emph {et~al.}(2008)\citenamefont {Ando},
  \citenamefont {Takahashi}, \citenamefont {Harii}, \citenamefont {Sasage},
  \citenamefont {Ieda}, \citenamefont {Maekawa},\ and\ \citenamefont
  {Saitoh}}]{Ando2008}%
  \BibitemOpen
  \bibfield  {author} {\bibinfo {author} {\bibfnamefont {K.}~\bibnamefont
  {Ando}}, \bibinfo {author} {\bibfnamefont {S.}~\bibnamefont {Takahashi}},
  \bibinfo {author} {\bibfnamefont {K.}~\bibnamefont {Harii}}, \bibinfo
  {author} {\bibfnamefont {K.}~\bibnamefont {Sasage}}, \bibinfo {author}
  {\bibfnamefont {J.}~\bibnamefont {Ieda}}, \bibinfo {author} {\bibfnamefont
  {S.}~\bibnamefont {Maekawa}},\ and\ \bibinfo {author} {\bibfnamefont
  {E.}~\bibnamefont {Saitoh}},\ }\bibfield  {title} {\bibinfo {title} {Electric
  manipulation of spin relaxation using the spin hall effect},\ }\href
  {https://doi.org/10.1103/PhysRevLett.101.036601} {\bibfield  {journal}
  {\bibinfo  {journal} {Phys. Rev. Lett.}\ }\textbf {\bibinfo {volume} {101}},\
  \bibinfo {pages} {036601} (\bibinfo {year} {2008})}\BibitemShut {NoStop}%
\bibitem [{\citenamefont {Czeschka}\ \emph {et~al.}(2011)\citenamefont
  {Czeschka}, \citenamefont {Dreher}, \citenamefont {Brandt}, \citenamefont
  {Weiler}, \citenamefont {Althammer}, \citenamefont {Imort}, \citenamefont
  {Reiss}, \citenamefont {Thomas}, \citenamefont {Schoch}, \citenamefont
  {Limmer}, \citenamefont {Huebl}, \citenamefont {Gross},\ and\ \citenamefont
  {Goennenwein}}]{Czeschka2011}%
  \BibitemOpen
  \bibfield  {author} {\bibinfo {author} {\bibfnamefont {F.~D.}\ \bibnamefont
  {Czeschka}}, \bibinfo {author} {\bibfnamefont {L.}~\bibnamefont {Dreher}},
  \bibinfo {author} {\bibfnamefont {M.~S.}\ \bibnamefont {Brandt}}, \bibinfo
  {author} {\bibfnamefont {M.}~\bibnamefont {Weiler}}, \bibinfo {author}
  {\bibfnamefont {M.}~\bibnamefont {Althammer}}, \bibinfo {author}
  {\bibfnamefont {I.-M.}\ \bibnamefont {Imort}}, \bibinfo {author}
  {\bibfnamefont {G.}~\bibnamefont {Reiss}}, \bibinfo {author} {\bibfnamefont
  {A.}~\bibnamefont {Thomas}}, \bibinfo {author} {\bibfnamefont
  {W.}~\bibnamefont {Schoch}}, \bibinfo {author} {\bibfnamefont
  {W.}~\bibnamefont {Limmer}}, \bibinfo {author} {\bibfnamefont
  {H.}~\bibnamefont {Huebl}}, \bibinfo {author} {\bibfnamefont
  {R.}~\bibnamefont {Gross}},\ and\ \bibinfo {author} {\bibfnamefont
  {S.~T.~B.}\ \bibnamefont {Goennenwein}},\ }\bibfield  {title} {\bibinfo
  {title} {Scaling behavior of the spin pumping effect in ferromagnet-platinum
  bilayers},\ }\href {https://doi.org/10.1103/PhysRevLett.107.046601}
  {\bibfield  {journal} {\bibinfo  {journal} {Phys. Rev. Lett.}\ }\textbf
  {\bibinfo {volume} {107}},\ \bibinfo {pages} {046601} (\bibinfo {year}
  {2011})}\BibitemShut {NoStop}%
\bibitem [{\citenamefont {Morota}\ \emph {et~al.}(2011)\citenamefont {Morota},
  \citenamefont {Niimi}, \citenamefont {Ohnishi}, \citenamefont {Wei},
  \citenamefont {Tanaka}, \citenamefont {Kontani}, \citenamefont {Kimura},\
  and\ \citenamefont {Otani}}]{Morota2011}%
  \BibitemOpen
  \bibfield  {author} {\bibinfo {author} {\bibfnamefont {M.}~\bibnamefont
  {Morota}}, \bibinfo {author} {\bibfnamefont {Y.}~\bibnamefont {Niimi}},
  \bibinfo {author} {\bibfnamefont {K.}~\bibnamefont {Ohnishi}}, \bibinfo
  {author} {\bibfnamefont {D.~H.}\ \bibnamefont {Wei}}, \bibinfo {author}
  {\bibfnamefont {T.}~\bibnamefont {Tanaka}}, \bibinfo {author} {\bibfnamefont
  {H.}~\bibnamefont {Kontani}}, \bibinfo {author} {\bibfnamefont
  {T.}~\bibnamefont {Kimura}},\ and\ \bibinfo {author} {\bibfnamefont
  {Y.}~\bibnamefont {Otani}},\ }\bibfield  {title} {\bibinfo {title}
  {Indication of intrinsic spin hall effect in $4d$ and $5d$ transition
  metals},\ }\href {https://doi.org/10.1103/PhysRevB.83.174405} {\bibfield
  {journal} {\bibinfo  {journal} {Phys. Rev. B}\ }\textbf {\bibinfo {volume}
  {83}},\ \bibinfo {pages} {174405} (\bibinfo {year} {2011})}\BibitemShut
  {NoStop}%
\bibitem [{SM()}]{SM}%
  \BibitemOpen
  \href@noop {} {}\bibinfo {note} {{See Supplemental Material for details on
  the magnetization angle dependence of the magnon-phonon hybrid dispersion,
  helicity-dependent phonon absorption, and phonon-driven magnetization
  precession, as well as the electrical detection scheme for spin
  current.}}\BibitemShut {Stop}%
\bibitem [{\citenamefont {Eyrich}\ \emph {et~al.}(2012)\citenamefont {Eyrich},
  \citenamefont {Huttema}, \citenamefont {Arora}, \citenamefont {Montoya},
  \citenamefont {Rashidi}, \citenamefont {Burrowes}, \citenamefont {Kardasz},
  \citenamefont {Girt}, \citenamefont {Heinrich}, \citenamefont {Mryasov},
  \citenamefont {From},\ and\ \citenamefont {Karis}}]{Eyrich2012}%
  \BibitemOpen
  \bibfield  {author} {\bibinfo {author} {\bibfnamefont {C.}~\bibnamefont
  {Eyrich}}, \bibinfo {author} {\bibfnamefont {W.}~\bibnamefont {Huttema}},
  \bibinfo {author} {\bibfnamefont {M.}~\bibnamefont {Arora}}, \bibinfo
  {author} {\bibfnamefont {E.}~\bibnamefont {Montoya}}, \bibinfo {author}
  {\bibfnamefont {F.}~\bibnamefont {Rashidi}}, \bibinfo {author} {\bibfnamefont
  {C.}~\bibnamefont {Burrowes}}, \bibinfo {author} {\bibfnamefont
  {B.}~\bibnamefont {Kardasz}}, \bibinfo {author} {\bibfnamefont
  {E.}~\bibnamefont {Girt}}, \bibinfo {author} {\bibfnamefont {B.}~\bibnamefont
  {Heinrich}}, \bibinfo {author} {\bibfnamefont {O.~N.}\ \bibnamefont
  {Mryasov}}, \bibinfo {author} {\bibfnamefont {M.}~\bibnamefont {From}},\ and\
  \bibinfo {author} {\bibfnamefont {O.}~\bibnamefont {Karis}},\ }\bibfield
  {title} {\bibinfo {title} {Exchange stiffness in thin film {Co} alloys},\
  }\href {https://doi.org/10.1063/1.3679433} {\bibfield  {journal} {\bibinfo
  {journal} {J. Appl. Phys.}\ }\textbf {\bibinfo {volume} {111}},\ \bibinfo
  {pages} {07C919} (\bibinfo {year} {2012})}\BibitemShut {NoStop}%
\bibitem [{\citenamefont {McSkimin}(1955)}]{Mcskimin1955}%
  \BibitemOpen
  \bibfield  {author} {\bibinfo {author} {\bibfnamefont {H.~J.}\ \bibnamefont
  {McSkimin}},\ }\bibfield  {title} {\bibinfo {title} {Measurement of the
  {Elastic} {Constants} of {Single} {Crystal} {Cobalt}},\ }\href
  {https://doi.org/10.1063/1.1722007} {\bibfield  {journal} {\bibinfo
  {journal} {J. Appl. Phys.}\ }\textbf {\bibinfo {volume} {26}},\ \bibinfo
  {pages} {406} (\bibinfo {year} {1955})}\BibitemShut {NoStop}%
\bibitem [{\citenamefont {Mihai~Miron}\ \emph {et~al.}(2010)\citenamefont
  {Mihai~Miron}, \citenamefont {Gaudin}, \citenamefont {Auffret}, \citenamefont
  {Rodmacq}, \citenamefont {Schuhl}, \citenamefont {Pizzini}, \citenamefont
  {Vogel},\ and\ \citenamefont {Gambardella}}]{Miron2010}%
  \BibitemOpen
  \bibfield  {author} {\bibinfo {author} {\bibfnamefont {I.}~\bibnamefont
  {Mihai~Miron}}, \bibinfo {author} {\bibfnamefont {G.}~\bibnamefont {Gaudin}},
  \bibinfo {author} {\bibfnamefont {S.}~\bibnamefont {Auffret}}, \bibinfo
  {author} {\bibfnamefont {B.}~\bibnamefont {Rodmacq}}, \bibinfo {author}
  {\bibfnamefont {A.}~\bibnamefont {Schuhl}}, \bibinfo {author} {\bibfnamefont
  {S.}~\bibnamefont {Pizzini}}, \bibinfo {author} {\bibfnamefont
  {J.}~\bibnamefont {Vogel}},\ and\ \bibinfo {author} {\bibfnamefont
  {P.}~\bibnamefont {Gambardella}},\ }\bibfield  {title} {\bibinfo {title}
  {Current-driven spin torque induced by the {Rashba} effect in a ferromagnetic
  metal layer},\ }\href {https://doi.org/10.1038/nmat2613} {\bibfield
  {journal} {\bibinfo  {journal} {Nat. Mater.}\ }\textbf {\bibinfo {volume}
  {9}},\ \bibinfo {pages} {230} (\bibinfo {year} {2010})}\BibitemShut {NoStop}%
\bibitem [{\citenamefont {Brataas}\ \emph {et~al.}(2000)\citenamefont
  {Brataas}, \citenamefont {Nazarov},\ and\ \citenamefont
  {Bauer}}]{Brataas2000}%
  \BibitemOpen
  \bibfield  {author} {\bibinfo {author} {\bibfnamefont {A.}~\bibnamefont
  {Brataas}}, \bibinfo {author} {\bibfnamefont {Y.~V.}\ \bibnamefont
  {Nazarov}},\ and\ \bibinfo {author} {\bibfnamefont {G.~E.~W.}\ \bibnamefont
  {Bauer}},\ }\bibfield  {title} {\bibinfo {title} {Finite-element theory of
  transport in ferromagnet--normal metal systems},\ }\href
  {https://doi.org/10.1103/PhysRevLett.84.2481} {\bibfield  {journal} {\bibinfo
   {journal} {Phys. Rev. Lett.}\ }\textbf {\bibinfo {volume} {84}},\ \bibinfo
  {pages} {2481} (\bibinfo {year} {2000})}\BibitemShut {NoStop}%
\bibitem [{\citenamefont {Taga}\ \emph {et~al.}(2025)\citenamefont {Taga},
  \citenamefont {Hisatomi}, \citenamefont {Sasaki}, \citenamefont {Komiyama},
  \citenamefont {Matsumoto}, \citenamefont {Narita}, \citenamefont {Karube},
  \citenamefont {Shiota},\ and\ \citenamefont {Ono}}]{Taga2025}%
  \BibitemOpen
  \bibfield  {author} {\bibinfo {author} {\bibfnamefont {K.}~\bibnamefont
  {Taga}}, \bibinfo {author} {\bibfnamefont {R.}~\bibnamefont {Hisatomi}},
  \bibinfo {author} {\bibfnamefont {R.}~\bibnamefont {Sasaki}}, \bibinfo
  {author} {\bibfnamefont {H.}~\bibnamefont {Komiyama}}, \bibinfo {author}
  {\bibfnamefont {H.}~\bibnamefont {Matsumoto}}, \bibinfo {author}
  {\bibfnamefont {H.}~\bibnamefont {Narita}}, \bibinfo {author} {\bibfnamefont
  {S.}~\bibnamefont {Karube}}, \bibinfo {author} {\bibfnamefont
  {Y.}~\bibnamefont {Shiota}},\ and\ \bibinfo {author} {\bibfnamefont
  {T.}~\bibnamefont {Ono}},\ }\bibfield  {title} {\bibinfo {title} {Generation
  of {Phonons} with {Out}-of-{Plane} {Angular} {Momentum} by {Superposition} of
  {Longitudinal} {Surface} {Acoustic} {Phonons}},\ }\href
  {https://doi.org/10.3379/msjmag.2501R004} {\bibfield  {journal} {\bibinfo
  {journal} {J. Magn. Soc. Jpn.}\ }\textbf {\bibinfo {volume} {49}},\ \bibinfo
  {pages} {17} (\bibinfo {year} {2025})}\BibitemShut {NoStop}%
\bibitem [{\citenamefont {Zhu}\ \emph {et~al.}(2018)\citenamefont {Zhu},
  \citenamefont {Yi}, \citenamefont {Li}, \citenamefont {Xiao}, \citenamefont
  {Zhang}, \citenamefont {Yang}, \citenamefont {Kaindl}, \citenamefont {Li},
  \citenamefont {Wang},\ and\ \citenamefont {Zhang}}]{Zhu2018}%
  \BibitemOpen
  \bibfield  {author} {\bibinfo {author} {\bibfnamefont {H.}~\bibnamefont
  {Zhu}}, \bibinfo {author} {\bibfnamefont {J.}~\bibnamefont {Yi}}, \bibinfo
  {author} {\bibfnamefont {M.-Y.}\ \bibnamefont {Li}}, \bibinfo {author}
  {\bibfnamefont {J.}~\bibnamefont {Xiao}}, \bibinfo {author} {\bibfnamefont
  {L.}~\bibnamefont {Zhang}}, \bibinfo {author} {\bibfnamefont {C.-W.}\
  \bibnamefont {Yang}}, \bibinfo {author} {\bibfnamefont {R.~A.}\ \bibnamefont
  {Kaindl}}, \bibinfo {author} {\bibfnamefont {L.-J.}\ \bibnamefont {Li}},
  \bibinfo {author} {\bibfnamefont {Y.}~\bibnamefont {Wang}},\ and\ \bibinfo
  {author} {\bibfnamefont {X.}~\bibnamefont {Zhang}},\ }\bibfield  {title}
  {\bibinfo {title} {Observation of chiral phonons},\ }\href
  {https://doi.org/10.1126/science.aar2711} {\bibfield  {journal} {\bibinfo
  {journal} {Science}\ }\textbf {\bibinfo {volume} {359}},\ \bibinfo {pages}
  {579} (\bibinfo {year} {2018})}\BibitemShut {NoStop}%
\bibitem [{\citenamefont {Jeong}\ \emph {et~al.}(2022)\citenamefont {Jeong},
  \citenamefont {Kim}, \citenamefont {Seo}, \citenamefont {Park}, \citenamefont
  {Jeong}, \citenamefont {Kim}, \citenamefont {Lauter}, \citenamefont {Egami},
  \citenamefont {Han},\ and\ \citenamefont {Choi}}]{Jeong2022}%
  \BibitemOpen
  \bibfield  {author} {\bibinfo {author} {\bibfnamefont {S.~G.}\ \bibnamefont
  {Jeong}}, \bibinfo {author} {\bibfnamefont {J.}~\bibnamefont {Kim}}, \bibinfo
  {author} {\bibfnamefont {A.}~\bibnamefont {Seo}}, \bibinfo {author}
  {\bibfnamefont {S.}~\bibnamefont {Park}}, \bibinfo {author} {\bibfnamefont
  {H.~Y.}\ \bibnamefont {Jeong}}, \bibinfo {author} {\bibfnamefont {Y.-M.}\
  \bibnamefont {Kim}}, \bibinfo {author} {\bibfnamefont {V.}~\bibnamefont
  {Lauter}}, \bibinfo {author} {\bibfnamefont {T.}~\bibnamefont {Egami}},
  \bibinfo {author} {\bibfnamefont {J.~H.}\ \bibnamefont {Han}},\ and\ \bibinfo
  {author} {\bibfnamefont {W.~S.}\ \bibnamefont {Choi}},\ }\bibfield  {title}
  {\bibinfo {title} {Unconventional interlayer exchange coupling via chiral
  phonons in synthetic magnetic oxide heterostructures},\ }\href
  {https://doi.org/10.1126/sciadv.abm4005} {\bibfield  {journal} {\bibinfo
  {journal} {Sci. Adv.}\ }\textbf {\bibinfo {volume} {8}},\ \bibinfo {pages}
  {eabm4005} (\bibinfo {year} {2022})}\BibitemShut {NoStop}%
\bibitem [{\citenamefont {Ishito}\ \emph {et~al.}(2023)\citenamefont {Ishito},
  \citenamefont {Mao}, \citenamefont {Kousaka}, \citenamefont {Togawa},
  \citenamefont {Iwasaki}, \citenamefont {Zhang}, \citenamefont {Murakami},
  \citenamefont {Kishine},\ and\ \citenamefont {Satoh}}]{Ishito2023}%
  \BibitemOpen
  \bibfield  {author} {\bibinfo {author} {\bibfnamefont {K.}~\bibnamefont
  {Ishito}}, \bibinfo {author} {\bibfnamefont {H.}~\bibnamefont {Mao}},
  \bibinfo {author} {\bibfnamefont {Y.}~\bibnamefont {Kousaka}}, \bibinfo
  {author} {\bibfnamefont {Y.}~\bibnamefont {Togawa}}, \bibinfo {author}
  {\bibfnamefont {S.}~\bibnamefont {Iwasaki}}, \bibinfo {author} {\bibfnamefont
  {T.}~\bibnamefont {Zhang}}, \bibinfo {author} {\bibfnamefont
  {S.}~\bibnamefont {Murakami}}, \bibinfo {author} {\bibfnamefont {J.-i.}\
  \bibnamefont {Kishine}},\ and\ \bibinfo {author} {\bibfnamefont
  {T.}~\bibnamefont {Satoh}},\ }\bibfield  {title} {\bibinfo {title} {{Truly
  chiral phonons in $\alpha$-{HgS}}},\ }\href
  {https://doi.org/10.1038/s41567-022-01790-x} {\bibfield  {journal} {\bibinfo
  {journal} {Nat. Phys.}\ }\textbf {\bibinfo {volume} {19}},\ \bibinfo {pages}
  {35} (\bibinfo {year} {2023})}\BibitemShut {NoStop}%
\bibitem [{\citenamefont {Ohe}\ \emph {et~al.}(2024)\citenamefont {Ohe},
  \citenamefont {Shishido}, \citenamefont {Kato}, \citenamefont {Utsumi},
  \citenamefont {Matsuura},\ and\ \citenamefont {Togawa}}]{Ohe2024}%
  \BibitemOpen
  \bibfield  {author} {\bibinfo {author} {\bibfnamefont {K.}~\bibnamefont
  {Ohe}}, \bibinfo {author} {\bibfnamefont {H.}~\bibnamefont {Shishido}},
  \bibinfo {author} {\bibfnamefont {M.}~\bibnamefont {Kato}}, \bibinfo {author}
  {\bibfnamefont {S.}~\bibnamefont {Utsumi}}, \bibinfo {author} {\bibfnamefont
  {H.}~\bibnamefont {Matsuura}},\ and\ \bibinfo {author} {\bibfnamefont
  {Y.}~\bibnamefont {Togawa}},\ }\bibfield  {title} {\bibinfo {title}
  {Chirality-induced selectivity of phonon angular momenta in chiral quartz
  crystals},\ }\href {https://doi.org/10.1103/PhysRevLett.132.056302}
  {\bibfield  {journal} {\bibinfo  {journal} {Phys. Rev. Lett.}\ }\textbf
  {\bibinfo {volume} {132}},\ \bibinfo {pages} {056302} (\bibinfo {year}
  {2024})}\BibitemShut {NoStop}%
\bibitem [{\citenamefont {Kim}\ \emph {et~al.}()\citenamefont {Kim},
  \citenamefont {Hwang}, \citenamefont {Moon}, \citenamefont {An},
  \citenamefont {Lee}, \citenamefont {Ko}, \citenamefont {Park}, \citenamefont
  {Choi},\ and\ \citenamefont {Hwang}}]{Kim2025}%
  \BibitemOpen
  \bibfield  {author} {\bibinfo {author} {\bibfnamefont {C.}~\bibnamefont
  {Kim}}, \bibinfo {author} {\bibfnamefont {I.-K.}\ \bibnamefont {Hwang}},
  \bibinfo {author} {\bibfnamefont {K.-W.}\ \bibnamefont {Moon}}, \bibinfo
  {author} {\bibfnamefont {K.}~\bibnamefont {An}}, \bibinfo {author}
  {\bibfnamefont {K.-J.}\ \bibnamefont {Lee}}, \bibinfo {author} {\bibfnamefont
  {J.-H.}\ \bibnamefont {Ko}}, \bibinfo {author} {\bibfnamefont {B.-G.}\
  \bibnamefont {Park}}, \bibinfo {author} {\bibfnamefont {K.-Y.}\ \bibnamefont
  {Choi}},\ and\ \bibinfo {author} {\bibfnamefont {C.}~\bibnamefont {Hwang}},\
  }\bibfield  {title} {\bibinfo {title} {{Chiral Acoustic Phonon and
  Conservation of Pseudoangular Momentum in $\alpha$-Quartz}},\ }\href
  {https://doi.org/https://doi.org/10.1002/adma.202511289} {\bibfield
  {journal} {\bibinfo  {journal} {Adv. Mater.}\ }\textbf {\bibinfo {volume}
  {38}},\ \bibinfo {pages} {e11289}}\BibitemShut {NoStop}%
\end{thebibliography}%

\clearpage 
\onecolumngrid 

\begin{center}
\textbf{\large Supplemental Material: Helicity-Selective Phonon Absorption and Phonon-Induced Spin Torque from Interfacial Spin-Lattice Coupling}
\end{center}
\vspace{3mm}

\setcounter{equation}{0}
\setcounter{figure}{0}
\setcounter{table}{0}
\setcounter{page}{1}
\renewcommand{\theequation}{S\arabic{equation}}
\renewcommand{\thefigure}{S\arabic{figure}}


\section{Magnetization angle dependence}

Here we take the equilibrium magnetization to arbitrary angle, ${\v m}_0 = (\sin\theta\cos\phi, \sin\theta\sin\phi, \cos\theta)$,
and describe small transverse fluctuations as $\delta {\v m} \approx \v R(\theta)(m'_x,m'_y, 0)$ with $|m'_{x,y}| \ll 1$, where
$R(\theta)$ is a rotation matrix that aligns the local quantization axis $z'$ with the equilibrium direction.
Dropping the prime notation and denote the magnetization fluctuations simply by $m_x$ and $m_y$, we have
\begin{equation}
\begin{aligned}\label{SLeq}
{\cal L}_\text{SL} &= \lambda_\text{SL} \left(m_x \cos\theta{\dot u}_y  - m_y {\dot u}_x \right)\nn
&=-\frac{i \lambda_\text{SL}}{2}\Big\{\left[(1+\cos\theta)m_- - (1-\cos\theta)m_+\right] \dot u_+ e^{-i\phi}
- \left[(1+\cos\theta) m_+ - (1-\cos\theta) m_- \right]\dot u_-e^{i\phi}\Big\},
\end{aligned}
\end{equation}
where $m_\pm = m_x \pm i m_y$ and $\dot{u}_\pm = \dot{u}_x \pm i \dot{u}_y$ represent the circular components of the magnetization and lattice velocity, respectively.
Note that the azimuthal angle $\phi$ enters the expression only as a phase factor $e^{\pm i\phi}$, and thus does not affect the helicity-selectivity of the coupling.

Equation \eqref{SLeq} clearly shows that the helicity-helicity coupling survives only when a perpendicular magnetization component is present ($\theta \neq \pi/2$). For the case of purely in-plane magnetization ($\theta = \pi/2$), the Lagrangian simplifies to:
\begin{equation}
\begin{aligned}\label{SLeq}
{\cal L}_\text{SL} \left(\theta = \frac{\pi}{2}\right)
&=-\frac{\lambda_\text{SL}}{2}m_y \left(\dot u_+ e^{-i\phi} + \dot u_- e^{i\phi}\right),
\end{aligned}
\end{equation}
In this limit, the distinctive helicity-helicity coupling vanishes as the lattice dynamics couples to the linear $m_y$ component, which is helicity neutral.

\subsection{Magnon-phonon hybrid dispersion}

\begin{figure}[h]
\includegraphics[width=0.8\columnwidth]{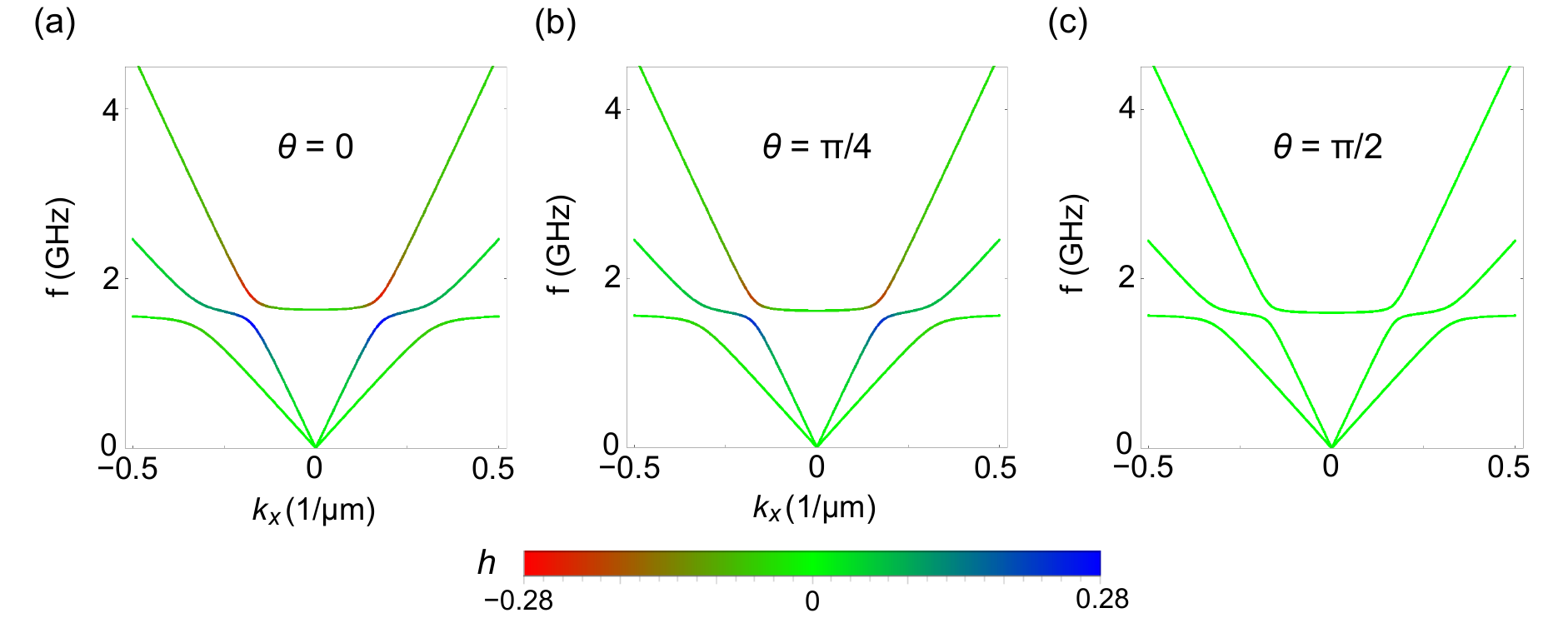}
\caption{\textbf{Magnon-phonon hybrid band structures.}
Hybridized dispersions in the presence of interfacial spin-lattice coupling for (a) $\theta = 0$, (b) $\theta = \pi/4$, and (c) $\theta = \pi/2$.
The color scale illustrates the phonon helicity weighted by the phonon energy ratio: $h = \frac{|u_+|^2 - |u_-|^2}{|u_+|^2 + |u_-|^2} \frac{E_\text{ph}}{E_\text{ph} + E_\text{mag}}$
, where $E_\text{ph}$ ($E_\text{mag}$) denotes the bare phonon (magnon) energy.}\label{fig:S1}
\end{figure}

Figure~\ref{fig:S1} shows the magnetization angle dependence of the magnon-phonon hybrid dispersion.
As clearly seen in Fig.~\ref{fig:S1}(c), the phonon helicity completely vanishes at $\theta = \pi/2$.
We also note that the band structure and phonon helicity profiles are independent of the azimuthal angle $\phi$.

\subsection{Helicity-dependent phonon absorption}

\begin{figure}[h]
\centering
\includegraphics[width=1.0\columnwidth]{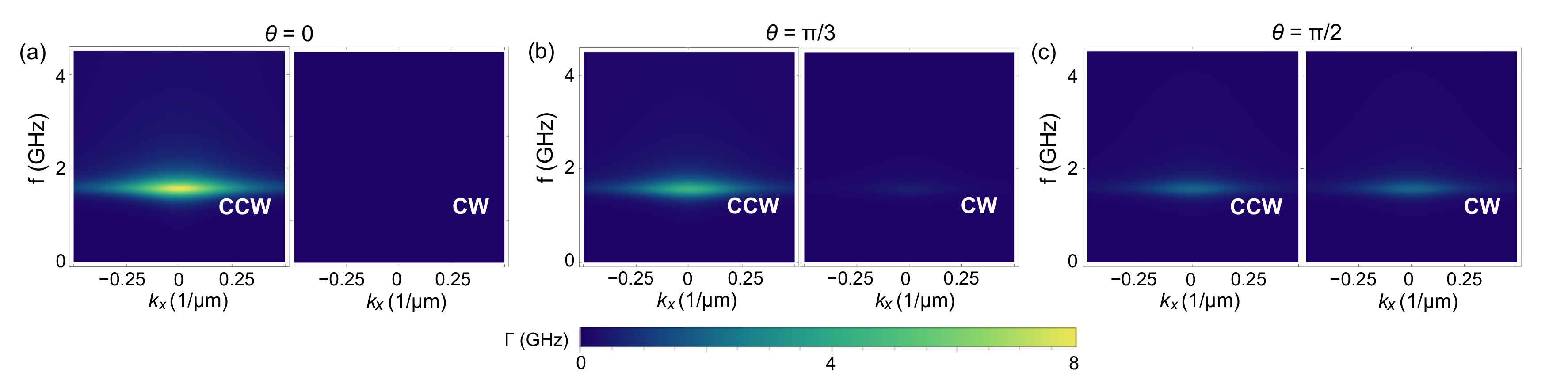}
\caption{\textbf{Helicity-dependent phonon absorption.}
Phonon absorption rates, $\Gamma_\text{SL}=\Delta P/\langle \mathcal{E}_\text{ph}\rangle$, for CCW and CW helicities at (a) $\theta = 0$, (b) $\theta = \pi/3$, and (c) $\theta = \pi/2$, where $\Delta P$ is the power transferred from phonons to magnons. The parameters used are $B_{\text{eff}}=0.056$~T and $\alpha=0.1$.
}\label{fig:S2}
\end{figure}

Figure~\ref{fig:S2} shows the magnetization angle dependence of the phonon absorption rate.
As shown in Fig.~\ref{fig:S2}(a), the absorption rate exhibits a stark contrast between counter-clockwise (CCW) and clockwise (CW) phonons for $\theta = 0$. However, at $\theta = \pi/2$, the absorption rates for both CCW and CW phonons become identical.
We also note that the absorption rate profiles are independent of the azimuthal angle $\phi$.

\subsection{Phonon-driven magnetization precession}

\begin{figure}[h]
\includegraphics[width=0.8\columnwidth]{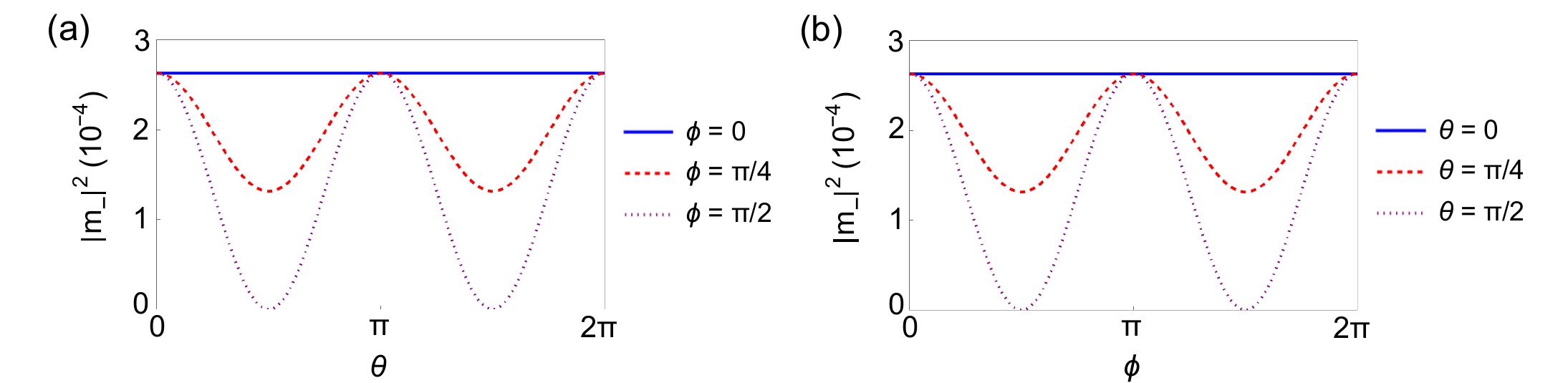}
\caption{\textbf{Phonon-driven magnetization precession.}
Angular dependence of the magnetization precession amplitudes on (a) $\theta$ and (b) $\phi$.
The parameters used are $B_{\text{eff}} = 0.056$~T, $\alpha = 0.1$ and $u_0 = 3.5$ pm.
}\label{fig:S3}
\end{figure}

In contrast to the phonon helicity and absorption profiles shown in Figs.~\ref{fig:S1} and \ref{fig:S2}, the magnetization precession driven by linearly polarized phonons depends on the azimuthal angle $\phi$. This dependence arises because $\phi$ determines the relative angle between the magnetization direction and the phonon polarization. Without loss of generality, we set the phonon polarization along the $x$-axis ($u_x = u_0\cos\omega t, u_y = 0$), as any in-plane rotation of the polarization direction can be equivalently described by a shift in the azimuthal angle $\phi$. Under this setting, the linear response of the magnetization fluctuations is given by:
\begin{equation}
\begin{aligned}\label{mpm}
m_\pm &= \mp \frac{\lambda_\text{SL} \omega}{\sqrt2 \rho_s({\omega \pm \tilde\Omega_{\v k} \mp i \alpha\omega})} \left(\cos\phi \mp i \cos\theta \sin\phi \right)u_0.
\end{aligned}
\end{equation}
The squared amplitude of these fluctuations follows the relation:
\begin{equation}
|m_\pm |^2 \propto 1 - \sin^2\theta \sin^2\phi.
\end{equation}
The expression $1 - \sin^2\theta \sin^2\phi$ is symmetric under the exchange of $\theta$ and $\phi$. This implies that the magnetization precession exhibits an identical functional dependence on both the polar and azimuthal angles, as shown in Fig.~\ref{fig:S3}.
Notably, the spin-current amplitude maintains a constant maximum value when the magnetization lies in the $xz$-plane $(\phi = 0)$.

\section{Electrical detection of spin currents}

\begin{figure}[h]
\includegraphics[width=0.4\columnwidth]{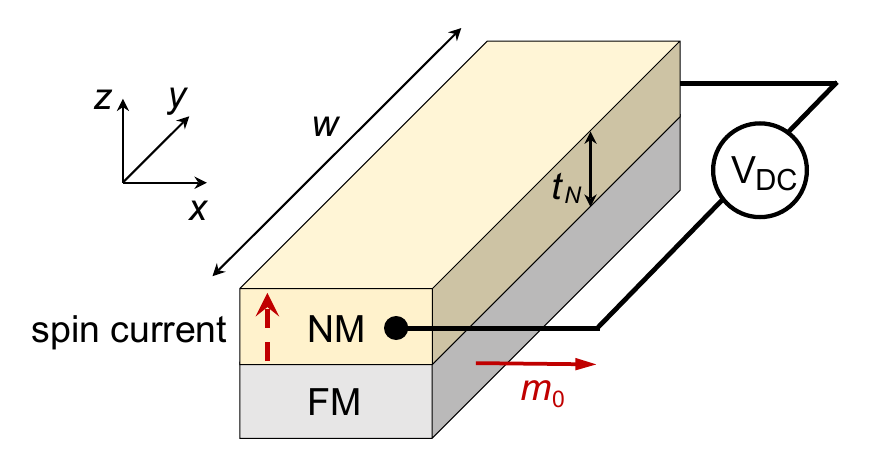}
\caption{\textbf{Device geometries for electrical spin detection via ISHE.}
Schematic view of geometry of ISHE detection scheme.
}\label{fig:S4}
\end{figure}

The spin current generated by phonons can be converted into an experimentally measurable voltage through the inverse spin Hall effect (ISHE).
For ISHE-based detection, the device typically consists of an NM/FM bilayer as shown in Fig.~\ref{fig:S4}.
Within a one-dimensional diffusion model, the spin accumulation $\mu_s(z) \equiv \mu_\uparrow(z) - \mu_\downarrow(z)$ obeys the diffusion equation $\partial_z^2 \mu_s(z) = \mu_s(z) / \lambda_N^2$, where $z$ denotes the spin-current transport direction and $\lambda_N$ is the spin-diffusion length of the NM. The corresponding spin-current density is given by $j_s(z) = -(\sigma_N/2e) \partial_z \mu_s(z)$, with $\sigma_N$ being the electrical conductivity of the NM.
In this geometry, physical termination of the NM layer imposes a vanishing spin-current condition, $j_s(t_N) = 0$.
At the injection interface ($z=0$), the boundary condition is given by~\cite{Tserkovnyak2002prb, Tserkovnyak2005}
\begin{equation}\label{bdz0}
j_s(0) = j_s^{\text{pump}} - j_s^\text{back},
\end{equation}
where the spin backflow is
\begin{equation}
j_s^{\text{back}} = \frac{e}{h} \frac{\text{Re}[g_{\uparrow\downarrow}]}{4\pi} \mu_s(0).
\end{equation}
Applying these boundary conditions yields the spin-current density profile within the NM layer:
\begin{equation}\label{jsz}
j_s(z) = \frac{j_s^{\text{pump} } \sinh[(z - t_N)/\lambda_N]}{\sinh(t_N/\lambda_N) + \epsilon \cosh(t_N/\lambda_N)},
\end{equation}
where
\begin{equation}
\epsilon =\frac{e^2}{h} \frac{\text{Re}[g_{\uparrow\downarrow}]}{4\pi} \frac{2\lambda_N}{\sigma_N}
\end{equation}
is the dimensionless correction factor accounting for the spin backflow.

Because the spin polarization is parallel to the magnetization direction ${\v m}^0$,
the ISHE converts spin current flowing along the $z$-axis into a transverse charge current ${\v j}^c \propto {\v j}_s \times {\v m}^0$.
Specifically, the $y$-directional charge current density $j^c_y$, selectively probes the spin polarization component parallel to the $x$-axis:
\begin{equation}
j^c_y(z) = -\theta_\text{SH} j_s(z) m^0_x.
\end{equation}
Using $E_y(z) = j^c_y(z)/\sigma$ and averaging over thickness $t_N$, we have
\begin{equation}
V = - \frac{w}{t_N} \int_0^{t_N} dz E_y(z)
\end{equation}

To quantitatively estimate the expected voltage signals, we consider a standard material system comprising a Pt/FM bilayer with the following representative parameters:
$\sigma_\text{Pt} = 6.4 \times 10^6 \, \Omega^{-1}\text{m}^{-1}$, $\theta_\text{Pt} = 0.08$, and $\lambda_\text{Pt} =  7 \, \text{nm}$~\cite{Ando2008}.
Assuming an NM thickness of $t_N = 10 \, \text{nm}$, a device width of $w = 1 \, \text{mm}$, and a pumped spin-current density of $j_s^\text{pump} = 64\, \text{A/cm}^2$
polarized along the $x$-direction (consistent with the main text), the estimated ISHE voltage is $V \approx 2.5 \, \mu\text{V}$.

\end{document}